\documentclass[twocolumn]{aastex631}

\newcommand\aastex{AAS\TeX}

\usepackage{xcolor}
\usepackage{wrapfig}
\usepackage{url}
\usepackage{lineno}
\usepackage{longtable}

\let\tablenum\relax
\usepackage{siunitx}

\received{\today}

\shorttitle{\aastex\ Kepler-102 Planet Masses and Compositions}
\shortauthors{Brinkman et al.}

\begin{document}

\title{Kepler-102: Masses and Compositions for a Super-Earth and Sub-Neptune Orbiting an Active Star}

\correspondingauthor{Casey L. Brinkman}
\email{clbrinkm@hawaii.edu}

\author[0000-0002-4480-310X]{Casey L. Brinkman}
\affiliation{Institute for Astronomy, University of Hawai'i, 2680 Woodlawn Drive, Honolulu, HI 96822 USA}

\author[0000-0002-3200-3121]{James Cadman}
\affiliation{SUPA, Institute for Astronomy, Royal Observatory, University of Edinburgh, Blackford Hill, Edinburgh EH93HJ, UK}
\affiliation{Centre for Exoplanet Science, University of Edinburgh, Edinburgh, UK}

\author[0000-0002-3725-3058]{Lauren Weiss}
\affiliation{Department of Physics, University of Notre Dame, In 46556}

\author[0000-0002-5258-6846]{Eric Gaidos}
\affiliation{Department of Earth Sciences, University of Hawai'i at M\"{a}noa, Honolulu, HI 96822 USA}

\author[0000-0002-6379-9185]{Ken Rice}
\affiliation{SUPA, Institute for Astronomy, Royal Observatory, University of Edinburgh, Blackford Hill, Edinburgh EH93HJ, UK}
\affiliation{Centre for Exoplanet Science, University of Edinburgh, Edinburgh, UK}

\author[0000-0001-8832-4488]{Daniel Huber}
\affiliation{Institute for Astronomy, University of Hawai'i, 2680 Woodlawn Drive, Honolulu, HI 96822 USA}

\author[0000-0001-8638-0320]{Zachary R. Claytor}
\affiliation{Institute for Astronomy, University of Hawai'i, 2680 Woodlawn Drive, Honolulu, HI 96822 USA}

\author{Aldo S. Bonomo}
\affiliation{INAF – Osservatorio Astrofisico di Torino, via Osservatorio 20, I-10025 Pino Torinese, Italy}

\author[0000-0003-1605-5666]{Lars A. Buchhave}
\affiliation{DTU Space, National Space Institute, Technical University of Denmark, Elektrovej 328, DK-
2800 Kgs. Lyngby, Denmark}

\author{Andrew Collier Cameron}
\affiliation{SUPA School of Physics and Astronomy, University of St Andrews, North Haugh, St
Andrews KY16 9SS, UK}

\author{Rosario Cosentino}
\affiliation{Fundación Galileo Galilei – INAF, Rambla J. A. F. Perez, 7, E-38712 S.C. Tenerife,
Spain}

\author[0000-0002-9332-2011]{Xavier Dumusque}
\affiliation{Observatoire Astronomique de l’Université de Genève, Chemin Pegasi 51b, CH-1290 Versoix, Switzerland}

\author{Aldo F Martinez Fiorenzano}
\affiliation{Fundación Galileo Galilei – INAF, Rambla J. A. F. Perez, 7, E-38712 S.C. Tenerife,
Spain}

\author[0000-0003-4702-5152]{Adriano Ghedina}
\affiliation{Fundación Galileo Galilei – INAF, Rambla J. A. F. Perez, 7, E-38712 S.C. Tenerife,
Spain}

\author{Avet Harutyunyan}
\affiliation{Fundación Galileo Galilei – INAF, Rambla J. A. F. Perez, 7, E-38712 S.C. Tenerife,
Spain}

\author[0000-0001-8638-0320]{Andrew Howard}
\affiliation{Department of Astronomy, California Institute of Technology, Pasadena, CA 91125, USA}

\author[0000-0002-0531-1073]{Howard Isaacson}
\affiliation{501 Campbell Hall, University of California at Berkeley, Berkeley, CA 94720, USA}

\author{David W. Latham}
\affiliation{Center for Astrophysics ${\rm \mid}$ Harvard {\rm \&} Smithsonian, 60 Garden Street, Cambridge, MA 02138, USA}

\author{Mercedes López-Morales}
\affiliation{Center for Astrophysics ${\rm \mid}$ Harvard {\rm \&} Smithsonian, 60 Garden Street, Cambridge, MA 02138, USA}

\author{Luca Malavolta}
\affiliation{Department of Physics and Astronomy, Università degli Studi di Padova, Vicolo dell’Osservatorio 3, IT-35122 Padova, Italy}
\affiliation{INAF – Osservatorio Astronomico di Padova, Vicolo dell’Osservatorio 5, IT-35122 Padova, Italy}

\author{Giuseppina Micela}
\affiliation{INAF – Osservatorio Astrofisico di Torino, via Osservatorio 20, I-10025 Pino Torinese,
Italy}

\author{Emilio Molinari}
\affiliation{INAF - Osservatorio Astronomico di Cagliari,
via della Scienza 5 , I-09047 Selargius, Italy}

\author{Francesco Pepe}
\affiliation{Observatoire Astronomique de l’Université de Genève, Chemin Pegasi 51b, CH-1290 Versoix, Switzerland}

\author{David F Philips}
\affiliation{Center for Astrophysics ${\rm \mid}$ Harvard {\rm \&} Smithsonian, 60 Garden Street, Cambridge, MA 02138, USA}

\author[0000-0003-1200-0473]{Ennio Poretti}
\affiliation{Fundación Galileo Galilei – INAF, Rambla J. A. F. Perez, 7, E-38712 S.C. Tenerife,
Spain}
\affiliation{INAF – Osservatorio Astronomico di Padova, Vicolo dell’Osservatorio 5, IT-35122 Padova,
Italy}

\author[0000-0002-7504-365X]{Alessandro Sozzetti}
\affiliation{INAF – Osservatorio Astrofisico di Torino, via Osservatorio 20, I-10025 Pino Torinese,
Italy}

\author{Stéphane Udry}
\affiliation{Observatoire Astronomique de l’Université de Genève, Chemin Pegasi 51b, CH-1290 Versoix, Switzerland}

\begin{abstract}

Radial velocity (RV) measurements of transiting multiplanet systems allow us to understand the densities and compositions of planets unlike those in the Solar System. Kepler-102, which consists of 5 tightly packed transiting planets, is a particularly interesting system since it includes a super-Earth (Kepler-102d) and a sub-Neptune-sized planet (Kepler-102e) for which masses can be measured using radial velocities. Previous work found a high density for Kepler-102d, suggesting a composition similar to that of Mercury, while Kepler-102e was found to have a density typical of sub-Neptune size planets; however, Kepler-102 is an active star, which can interfere with RV mass measurements. To better measure the mass of these two planets, we obtained 111 new RVs using Keck/HIRES and TNG/HARPS-N and modeled Kepler-102's activity using quasi-periodic Gaussian Process Regression. For Kepler-102d, we report a mass upper limit of M$_{d} < $5.3 M$_{\oplus}$ [95\% confidence], a best-fit mass of M$_{d}$=2.5 $\pm$ 1.4 M$_{\oplus}$, and a density of $\rho_{d}$=5.6 $\pm$ 3.2 g/cm$^{3}$ which is consistent with a rocky composition similar in density to the Earth. For Kepler-102e we report a mass of  M$_{e}$=4.7 $\pm$ 1.7 M$_{\oplus}$ and a density of $\rho_{e}$=1.8 $\pm$ 0.7 g/cm$^{3}$. These measurements suggest that Kepler-102e has a rocky core with a thick gaseous envelope comprising 2-4\% of the planet mass and 16-50\% of its radius. Our study is yet another demonstration that accounting for stellar activity in stars with clear rotation signals can yield more accurate planet masses, enabling a more realistic interpretation of planet interiors. 
\end{abstract}



\section{Introduction} 
\label{sec:intro}
The results of the \textit{Kepler} mission demonstrate that planets smaller than the size of Neptune form in abundance around stars like the Sun \citep{2012ApJS..201...15H, 2021AJ....161...36B}. Half of all Sun-like stars in the Galaxy host planets between the sizes of Earth and Neptune \citep{2013PNAS..11019273P}, but the sun does not. To understand the bulk composition of these planets, for which there is no Solar System analogue, requires precise density estimates and planetary interior modeling. 

Radial velocity (RV) measurements of transiting planets are a powerful tool to better understand the densities and compositions of other worlds. From RVs we can measure the mass of a planet, and from the transit depth we can measure its radius. Combining these gives the  planet's bulk density. However, while we have many hundreds of density measurements for giant planets, and many for sub-Neptunes with radii between 1.5-4 R$_{\oplus}$, we have RV determined densities for only 42 super-Earths with radii between 1-1.5 R$_{\oplus}$ (NASA Exoplanet Archive queried 09/14/22, \cite{2013PASP..125..989A}). 

In addition to the large number of RVs needed to measure the semi-amplitudes produced by Earth-size planets, star spots and active regions can affect the spectrum over time and are capable of masking the $\sim$1 m/s semi-amplitudes produced by small, rocky planets. Star spots (and plage) can mimic the signature of an orbiting planet as the star rotates. If the variability of the star can be modeled along with RV signal of the planets, the actual semi-amplitude of the planet can be measured \citep{2011A&A...525A.140D}. The method of Gaussian Process (GP) Regression describes the correlation between data points through a covariance matrix and was introduced to the field of exoplanets in 2011 to model correlated noise from instrument systematics \citep{2012MNRAS.419.2683G}. GP analysis was subsequently used to model correlated stellar variability due to the rotation and evolution of star spots \citep{2014MNRAS.443.2517H}, enabling mass measurements of several small planets around active stars, including \textit{Kepler}-78b \citep{2016IAUFM..29A.208G} and K2-291b \citep{2019AJ....157..116K}.    

Another complicating factor is the degeneracy of inferring planet compositions from measured densities. We assume to first order that sub-Jovian differentiated planets are composed of an iron core, a silicate rock mantle layer, a water or ice layer, and an atmosphere likely composed of hydrogen and helium \citep{2007ApJ...656..545V}. Equation-of-state modeling for silicate rock, iron, and ice allow partial constraints on possible compositions. However, it is often possible to explain the bulk density of a planet with multiple combinations of rock, metal, water/ice, and atmosphere, making the problem of finding a unique solution for composition inherently degenerate using bulk density alone \citep{2010ApJ...712..974R}. In some cases, the size and proximity of a planet to its host star can be used to justify excluding ice and atmosphere from probable compositions, since these constituents are expected to escape \citep{2017ApJ...847...29O, 2017MNRAS.472..245L}. 

From the Solar System, we expect large planets to have extensive low molecular weight atmospheres, while smaller planets are composed primarily of rock and metal. The masses and radii of small exoplanets suggest a transition between primarily rocky and gas-enveloped planets at approximately 1.5 $R_{\oplus}$ \citep{2014ApJ...783L...6W, 2015ApJ...801...41R, 2017AJ....154..109F}, with planets smaller than 1.5 $R_{\oplus}$ often having compositions consistent with Earth-like iron-to-silicate ratios \citep{2015ApJ...800..135D}. However, existing super-Earth mass measurements indicate a wide diversity of densities among those planets---far more diverse than we observe for rocky planets in our own Solar System \citep{2014ApJS..210...20M, 2016ApJ...822...86M, 2019ApJ...883...79D}. These RV densities suggest the interior compositions of Earth-size planets could potentially vary from water-rich, to entirely rocky, to predominantly iron \citep{2019NatAs...3..416B}.

Above R=1.5$R_{\oplus}$, planets tend to retain substantial H/He envelopes that increase the radius while decreasing the density. Neptune is often called an ``ice giant" planet, and initially many sub-Neptune size exoplanets were also modeled as icy planets. However many studies now argue that they are alternatively composed of a solid core with a gaseous envelope \citep{2015ApJ...801...41R, 2017ApJ...847...29O}, while others argue they are water worlds \citep{2019PNAS..116.9723Z}. Some recent studies have even suggested deep magma oceans for many sub-Neptunes discovered by \textit{Kepler}, which we expect to greatly impact the composition of the atmosphere \citep{2020ApJ...891..111K}. 

An interesting System for exploring planet compositions is Kepler-102, which consists of 5 tightly-packed transiting planets (within 0.16 au of their host star), including a super-Earth and a sub-Neptune size planet. Previously, masses have been measured for the super-Earth Kepler-102d (M$_{d}$=3.8 $\pm$ 1.8 M$_\oplus$) and sub-Neptune Kepler-102e (M$_{e}$=8.93 $\pm$ 2.0 M$_\oplus$) \citep{2014ApJS..210...20M}. The density of Kepler-102d was found to be $\rho_{e}$=13.3 $\pm$ 6.46 g/cm$^3$, suggesting the planet was consistent with a composition of pure compressed iron. Kepler-102e was found to have a density consistent with other sub-Neptunes. \cite{2014ApJS..210...20M} did not report masses for the other three planets, which are all smaller than the Earth. 

The radius of Kepler-102 was recently revised by \cite{2018PASP..130d4504F} using the \textit{Gaia} parallax, yielding more accurate and precise measurements for the radii of all transiting planets in the system. The density of Kepler-102d dropped from $\rho_{d}$=13.3 to $\rho_{d}$=9.5 g/cm$^3$, allowing the planet to have a mixed composition of iron and rock. While this revised density is more realistic, it is still high for a planet of its size, implying a Mercury-like composition. 

Kepler-102, however, is an active star as indicated by a Mt. Wilson S-Value of 0.41 \citep{2013A&A...554A..50S} and logR'HK of -4.6, and such activity could potentially interfere with the previous mass measurements for Kepler-102d and Kepler-102e. In this work, we present new mass and density measurements for these planets, using new RVs from Keck/HIRES and TNG/HARPS-N and applying Gaussian Process Regression to model the stellar activity in both the photometry and the RVs. We then explore new composition estimates using equation-of-state modeling. 

\section{Stellar Activity $\&$ Rotation}
\label{sec:stellar}
\subsection{\textit{Kepler} Photometry}

The \textit{Kepler} spacecraft, launched in 2009, observed one $>$100 square degree patch of the sky continuously until 2013, looking for the signatures of transiting exoplanets \citep{2010Sci...327..977B}. Photometry of Kepler-102 was gathered at a 30-minute cadence for 17 quarters of observations. The properties of Kepler-102 are given in Table \ref{table:stellar}.

\begin{figure*}

\includegraphics[width=1.0\textwidth]{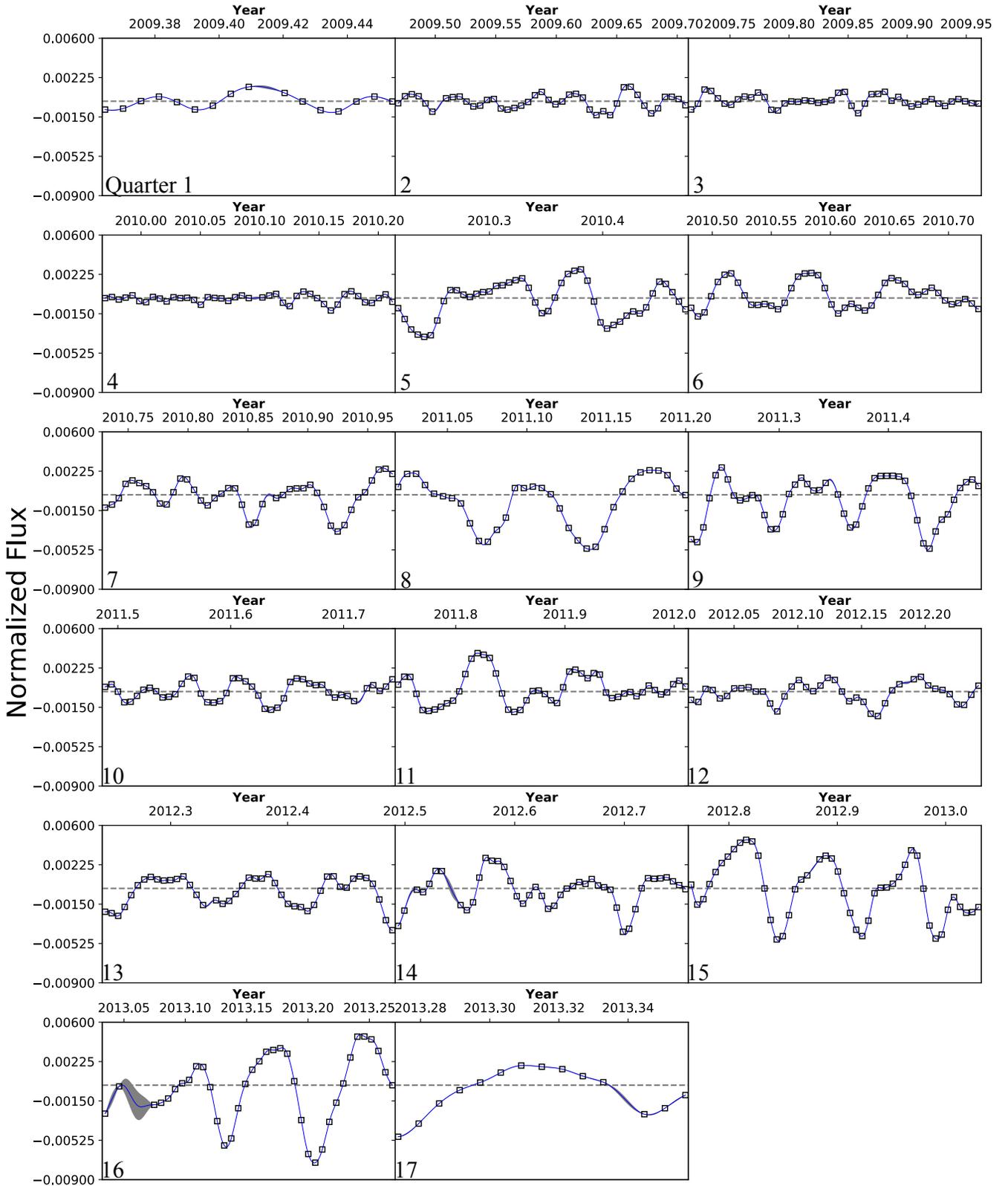}
\caption{PDCSAP Flux of Kepler-102 over 17 quarters of observation, binned every 10 hours (open boxes). Each panel corresponds to one quarter of photometry. The blue curve shows the Gaussian process regression of the photometry with the best-fit kernel hyperparameters from Table \ref{table:hyperparams}. The grey shaded regions show the uncertainty in the GP model.}
\label{fig:photom}
\end{figure*}

To assess whether the photometry of Kepler-102 showed the quasi-periodic rotational modulation characteristic of active stars, we downloaded the \textit{Kepler} Pre-search Data Conditioning SAP (PDCSAP) flux photometry using the package \texttt{Lightkurve} \citep{2018ascl.soft12013L}. In our analysis we used the full 17 quarters of photometry on Kepler-102, and we masked out the individual transits of Kepler-102d and Kepler-102e. The transit depths for the other three planets were negligible. We furthermore re-binned the photometry to have one data point every 10 hours to make our computations more manageable. While this removes information about stellar variability on timescales shorter than this, 10 hours is much less than the expected stellar rotation period and the star spot decay timescale. Figure \ref{fig:photom} shows the full \textit{Kepler} light curve over all 17 quarters. We observe a distinct rotation signal, consistent with the presence of star spots. 

We computed a rotation-based age for Kepler 102 using the stellar model grid fitting tools in \texttt{kiauhoku} \citep{2020ascl.soft11027C}. Using the magnetic rotational braking law of \cite{2013ApJ...776...67V}, we derived an age of 4.5 $\pm$ 1.1 Gyr. We note that using the stalled braking law of \cite{2016Natur.529..181V} did not significantly affect the age estimate. This is consistent with the Isochrone age from \cite{2018AJ....156..264F} to within 1$\sigma$. 

\begin{table}[h]
\begin{center}
    Kepler-102 Stellar Properties
\begin{tabular}{|c|c|c|}

\hline
Property                     & Value                & Source\\ \hline
Other Names                  & KOI 82               & B     \\ \hline
KIC ID                       & 10187017             & B     \\ \hline
RA (J2000)                   & 18 45 55.85          & B     \\ \hline
Dec (J2000)                  & +47 12 28.84         & B     \\ \hline
V(mag)                       & 12.07                & B     \\ \hline
$T_{\textrm{eff}}$(K)               & 4909 $\pm$ 98        & F     \\ \hline 
Metallicity [Fe/H]           & 0.11 $\pm$0.04       & F     \\ \hline 
Mass (M$_{\odot}$)           & 0.803$\pm$0.021      & F     \\ \hline
Radius (R$_{\odot}$)         & 0.724$\pm$0.018      & F     \\ \hline
Isochrone Age (Gyr)          & 1.07$_{-0.5}^{+3.6}$ & F     \\ \hline
Rotation-Based Age (Gyr)     & 4.5 $\pm$ 1.1        & *     \\ \hline
Trigonometric Parallax [mas] & 9.301 $\pm$ 0.019    & F     \\ \hline

\end{tabular}
\caption{ Stellar properties for Kepler-102 are compiled from the following sources: B=\cite{2011AJ....142..112B}, F=\cite{2018AJ....156..264F}, *=This Work}
\label{table:stellar}
\end{center}

\end{table}

\begin{figure*}
    \centering
    \includegraphics[width=0.8\textwidth]{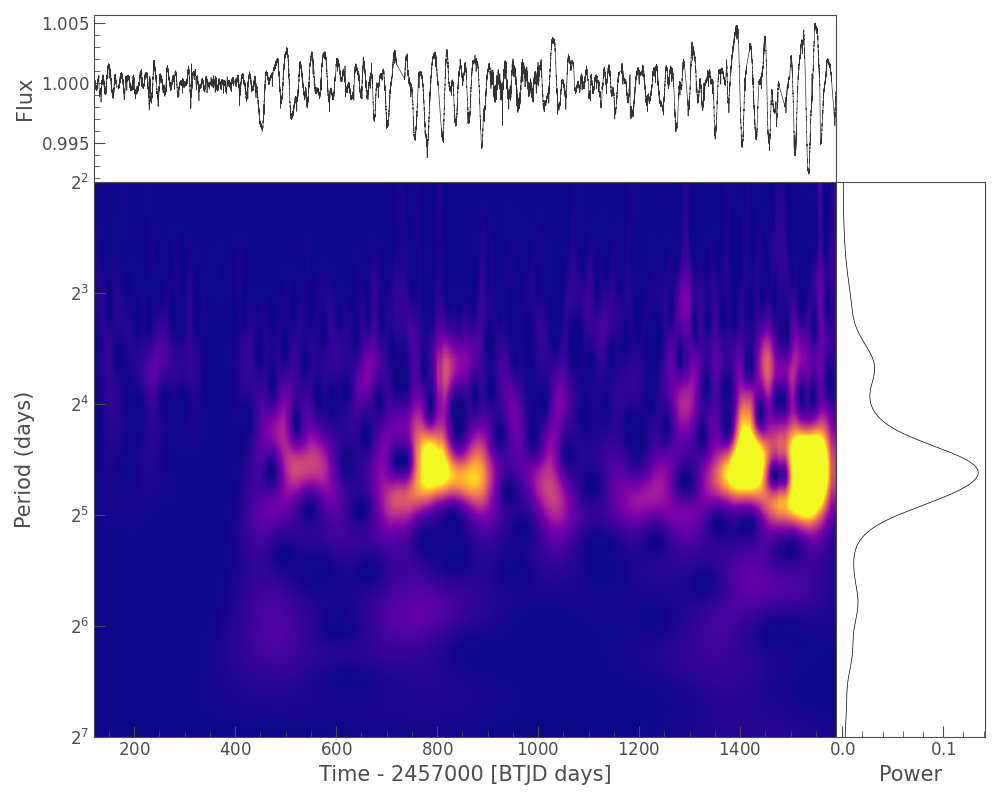}
    
    \caption{Wavelet analysis for the 4-year light curve of Kepler-102. The top panel shows the photometry over 17 quarters of \textit{Kepler} data, where time is measured in Barycentric TESS Julian Date. The central panel shows the time-variable frequency decomposition. The right panel shows the time-integrated wavelet power spectrum of the light curve. The clear period of maximum power is at 27 days, suggesting that this is the rotation period of the star, which we use to inform our GP kernel. There is also significant power from 10-32 days, showing that spots form, evolve, and disappear over time. The peak at 13 days is likely a half-period alias from spots appearing on opposite sides of the star \citep{2014MNRAS.443.2517H, 2018ApJ...863..190B}.}
    \label{fig:wavelet}
\end{figure*}

\begin{table*}[h]
\begin{center}
    Kepler-102 Planet Properties

\begin{tabular}{|l|l|l|l|l|}

\hline
Planet Name & Period (days)                 & T$_{c}$ (days)                 & Radius (R$_\oplus)$ & Mass (M$_\oplus)$         \\ \hline
Kepler-102b      & 5.28697 $\pm$0.00001   $^{G}$ & 2454968.870$\pm$ 0.001 $^{G}$  & 0.471 $\pm$ 0.024 $^{F}$  & --                  \\ \hline
Kepler-102c      & 7.07139$\pm$0.00002    $^{G}$ & 7.07139$\pm$ 0.00002 $^{G}$    & 0.554 $\pm$ 0.026 $^{F}$  & --                  \\ \hline
Kepler-102d      & 10.311767$\pm$0.000004 $^{G}$ & 2454967.0913$\pm$0.0001 $^{G}$ & 1.340 $\pm$ 0.089 $^{F}$  & 2.5 $\pm$ 1.4 $^{*}$\\ \hline
Kepler-102e      & 16.145699$\pm$0.000002 $^{G}$ & 2454967.7537$\pm$0.0001 $^{G}$ & 2.414 $\pm$ 0.142 $^{F}$  & 4.7 $\pm$ 1.7 $^{*}$\\ \hline
Kepler-102f      & 27.45359 $\pm$0.00006  $^{G}$ & 2454978.028$\pm$0.001 $^{G}$   & 0.753 $\pm$ 0.033 $^{F}$  & --                  \\ \hline
\end{tabular}
\caption{Parameters for the Kepler-102 system are compiled from the following sources: G=\cite{2019RAA....19...41G}, F=\cite{2018AJ....156..264F}, *=This Work. }
\label{table:planet}

\end{center}
\end{table*}

\subsection{Gaussian Process Model}

We use Gaussian Process Regression (hereafter GP) to characterize the stellar noise caused by the rotation of star spots and plage \citep{2006gpml.book.....R}. A commonly used model for this type of stellar noise is quasi-periodic: the rotation of the star produces a periodic signal, but evolution of star spots adds gradual phase offsets to the periodic signal. To capture this behavior we use the quasi-periodic kernel from the \texttt{RadVel} package \citep{2018PASP..130d4504F}, the product of a periodic function and a squared exponential function given as:
\begin{equation}
   C_{ij} =\eta_{1}^{2} \exp\left[-\frac{\mid t_{i}-t_{j}\mid^{2}}{\eta_{2}^{2}} -\frac{\sin^{2} (\mid t_{i}-t_{j}\mid/\eta_{3})}{2\eta_{4}^{2}}\right]
\end{equation}

where $\eta_{1}$ is the amplitude of the covariance, $\eta_{2}$ is the timescale of the exponential decay (here a proxy for the spots evolutionary timescale) , $\eta_{3}$ is the variability period (here the rotational period of the star), and $\eta_{4}$ is a coherence scale for the periodic variability. The terms $i$ and $j$ are the indices of the covariance matrix. 

GPs are non-parametric, meaning that they do not produce a predictive model of the data using fitting parameters. Instead, they use tunable hyper-parameters, which inform how the GP fits the covariance kernel to the data. In this analysis $\eta_{1}$, $\eta_{2}$, $\eta_{3}$, $\eta_{4}$ are hyper-parameters for the quasi-periodic kernel. 

The flexibility of GPs is a double-edged sword: GPs can be used to describe complex data sets that more traditional modeling cannot, but they also have ability to over-fit the data, resulting in the unintentional removal of planetary signals. One method of mitigating the over-fitting of data is to ensure that the GP has physically motivated priors for each hyper-parameter.

\subsection{Determining GP Priors}
\label{sec:training}
The rotation and evolution of star spots will produce similar signals in photometry and in the radial velocities. The rotational period of the star, the exponential decay timescale, and coherence scales should be consistent between photometry and RVs, while the amplitudes of each signal are in different units and should not be consistent. Therefore,  we can use the posterior results of the photometric fit as physically motivated starting values for $\eta_{2}$, $\eta_{3}$, and $\eta_{4}$. This should help the GP to distinguish between the stellar and planetary signals present in the RVs. The light curve is an ideal data set for this training because it is sampled at a much higher cadence than the sparse RV curve, which allows us to more accurately determine the hyper-parameters. 

Before fitting the light curve, we derived a starting value for the rotation period of the star by looking at the period of maximum power using a Wavelet transform (Figure \ref{fig:wavelet}). We used the Morlet wavelet transform from \cite{2020zndo...4100507V} in SciPy, and applied the power spectral density correction from \cite{2007JAtOT..24.2093L}. We observe a strong peak in the time-integrated power spectrum at $\sim$ 27 days, which we interpret as the rotation period of the star. Through the entire light curve, however, we can see a very pronounced evolution of the stellar activity through the change in periods that show significant power. In the first four quarters, the period of maximum power appears distinctly as $\sim$13 days without power at 27 days. This is indicative of spots on either side of the star, as they make the rotation period appear as half its value for the first four quarters, before one of the spots disappears \citep{2014MNRAS.443.2517H, 2018ApJ...866...99B}. 

First, we fit each of the seventeen quarters of photometry separately because as the light curve evolves, so can the hyper-parameters. Equation 1 is not sufficient to model a star that has spots with different decay timescales and rotation periods (e.g. latitudinal differential rotation), therefore fitting each quarter separately informs us of the variance in these timescales. The best-fit hyperparameters for each quarter are given in Table \ref{table:quarters}. We found that part of the light curve has a best-fit exponential decay length ($\eta_{2}$) of $\sim30-34$ days, while the other parts a best-fit value of $\sim22-28$ days. 

We then fit the full \textit{Kepler} photometry to determine the best overall hyper-parameters listed in Table \ref{table:hyperparams}. The best-fit $\eta_{1}$, $\eta_{3}$, and $\eta_{4}$ are similar to the median of the individual 17 quarters fits. The best fit $\eta_{2}$ (30 days) is not the same as the median of the 17 individual quarters (26 days), but falls within 1$\sigma$ of this value. We will use the hyper-parameters from the full light curve fit for the starting values in the RV fit, with a cautious treatment of $\eta_{2}$ as discussed later. Figure \ref{fig:photom} shows the photometry with the GP regression using the best-fit hyper-parameters listed in Table \ref{table:hyperparams}. 

We also used a Quasi-Periodic kernel to fit the \textit{Kepler} photometry in \texttt{PyORBIT} \citep{2016A&A...588A.118M, 2018AJ....155..107M}, a package for modelling planetary and stellar activity signals. The GP quasiperiodic kernel was applied through the \texttt{george} package \citep{2015ITPAM..38..252A}. Hyperparameter optimization was performed using the differential evolution code \texttt{pyDE}. The optimized hyperparameters were then given as priors to the affine-invarient ensemble sampler \texttt{emcee} \citep{2013PASP..125..306F}, from which we obtained our best-fit activity parameters. The best-fit values for the hyperparameters obtained from \texttt{} were found to agree well with those found using \texttt{RadVel}.

\begin{table}
\begin{center}
 GP Hyper-parameters by \textit{Kepler} Quarter

\begin{tabular}{|c|c|c|c|c|}

\hline
Q        & $\eta_{1}$        & $\eta_{2}$  & $\eta_{3}$  & $\eta_{4}$        \\ 
         & (Flux x 10$^{-3}$)  & (Days)      & (Days) &                        \\ \hline

1        & 1.0  $\pm$ 0.1    & 29 $\pm$ 3  & 26 $\pm$ 1  & 0.032 $\pm$ 0.06  \\ \hline
2        & 0.6  $\pm$ 0.08   & 28 $\pm$ 5  & 27 $\pm$ 1        & 0.34 $\pm$ 0.08   \\ \hline
3        & 0.5  $\pm$ 0.08   & 25 $\pm$ 4  & 26 $\pm$ 0.4      & 0.21 $\pm$ 0.01   \\ \hline
4        & 0.38 $\pm$ 0.08   & 23 $\pm$ 2  & 26 $\pm$ 0.5      & 0.13 $\pm$ 0.02   \\ \hline
5        & 1.5  $\pm$ 0.2    & 27 $\pm$ 2  & 26 $\pm$ 0.9      & 0.28 $\pm$ 0.02   \\ \hline
6        & 0.99 $\pm$ 0.2    & 34 $\pm$ 4  & 26 $\pm$ 0.2      & 0.22 $\pm$ 0.01   \\ \hline
7        & 1.2  $\pm$ 0.2    & 32 $\pm$ 5  & 27 $\pm$ 1        & 0.22 $\pm$ 0.02   \\ \hline
8        & 1.9  $\pm$ 0.4    & 31 $\pm$ 6  & 26  $\pm$ 0.3     & 0.29 $\pm$ 0.02   \\ \hline
9        & 1.6  $\pm$ 0.3    & 35 $\pm$ 5  & 26 $\pm$ 0.4      & 0.31 $\pm$ 0.03   \\ \hline
10       & 0.77 $\pm$ 0.6    & 22 $\pm$ 6  & 26 $\pm$ 1        & 0.26 $\pm$ 0.04   \\ \hline
11       & 1.3  $\pm$ 0.1    & 23 $\pm$ 3  & 26 $\pm$ 0.8      & 0.28 $\pm$ 0.01   \\ \hline
12       & 0.79 $\pm$ 0.1    & 23 $\pm$ 3  & 26 $\pm$ 0.9      & 0.16 $\pm$ 0.01   \\ \hline
13       & 1.2  $\pm$ 0.2    & 28 $\pm$ 4  & 26 $\pm$ 0.3      & 0.23 $\pm$ 0.02   \\ \hline
14       & 1.3  $\pm$ 0.2    & 24 $\pm$ 3  & 26 $\pm$ 0.9      & 0.16 $\pm$ 0.01   \\ \hline
15       & 2.0  $\pm$ 0.2    & 37 $\pm$ 4  & 26 $\pm$ 0.2      & 0.25 $\pm$ 0.01   \\ \hline
16       & 2.3  $\pm$ 0.4    & 34 $\pm$ 5  & 26 $\pm$ 1        & 0.18 $\pm$ 0.01   \\ \hline
17       & 2.0  $\pm$ 0.6    & 54 $\pm$ 20 & 26 $\pm$ 0.5      & 0.26 $\pm$ 0.2   \\ \hline

\end{tabular}
\caption{Best fit hyper-parameters for the Quasi-Periodic kernel are shown for each quarter of \textit{Kepler} photometry fit seperately. $\eta_{1}$ corresponds to the amplitude of covariance in normalized flux, $\eta_{2}$ to the exponential characteristic length, $\eta_{3}$ to the period of covariance (here, the rotation period of the star), and $\eta_{4}$ to the periodic coherence scale.}
\label{table:quarters}
\end{center}

\end{table}

\section{Planet Masses}
\label{sec:masses}
\subsection{Radial Velocity Data}
\subsubsection{HIRES}

We make use of 72 RVs from the HIRES echelle spectrometer on the 10 meter Keck I telescope \citep{1994SPIE.2198..362V}. \cite{2014ApJS..210...20M} obtained data between 2010-2013, totalling 35 RVs. The  California Planet Search (CPS) team collected 37 additional RVs spanning 2013 to 2019 (Weiss et al. in prep).

The spectrometer uses an iodine cell mounted in front of the slit in order to provide a benchmark spectrum against which the Doppler shift of Kepler-102 can be measured. The resulting shapes of the iodine lines account for changes in spectrometer optics by recording the PSF of the spectrometer \citep{1992PASP..104..270M}. We obtained a template spectrum by observing the star without the iodine cell. To model the PSF of the HIRES spectrograph we also observed rapidly-rotating B stars with the iodine cell in the light path immediately before and after the template. Each RV spectrum was then reproduced with the deconvolved template spectrum and a laboratory iodine atlas spectrum convolved with the HIRES PSF of the observation. Sky subtraction was performed through the use of a $14\arcsec.0$ long slit in order to observe enough sky compared to the Keck point-spread function (full-width half-max = $1\arcsec.0$) to get a robust background estimate. We used the standard CPS data reduction pipeline as described in \cite{2010ApJ...721.1467H}.

\subsubsection{HARPS-North}
We obtained a total of 78 RVs for Kepler-102 using the HARPS-N spectrograph installed on the 3.6m Telescopio Nazionale Galileo (TNG) at the Observatorio Roque de Los Muchachos in La Palma, Spain. These were collected as part of the HARPS-N Collaboration’s Guaranteed Time Observations (GTO) program, and span a period from June 2012 to July 2020. The RVs are listed in Table \ref{table:rvs}.

A CCD change after the observation at BJD 2456162.45 meant that we separated the 78 RVs into two distinct datasets in order to account for this change. We collected a total of 23 RVs prior to the CCD change and 55 afterwards. Each of these observations had an exposure time of 1800s. The spectra prior to the CCD change were reduced using the HARPS-N Data Reduction Software (DRS) 3.7. The spectra after the CCD change were reduced using the new DRS 2.2.8 \citep{2021A&A...648A.103D}.

We discovered anomalously large uncertainties in three of the RVs at BJD=2456102.48, 2456606.34, and 2456606.37, most likely due to poor seeing according to the observing logs, and we therefore excluded them from our analysis. With the exception of these, our remaining 22 observations prior to the CCD change had S/N values between 24.8 and 54.4 per extracted pixel, with an average value of 41.2 at 550nm and a mean RV precision of 2.14\,ms$^{-1}$. The remaining 53 RVs after the CCD change had S/N values between 19.72 and 67.12 per extracted pixel, with an average value of 49.63 at 550nm and a mean RV precision of 2.27\,ms$^{-1}$.

\begin{table}[]
\begin{center}

\begin{tabular}{|c|c|c|c|c|c|}
\hline
Time               & RV       & RV unc  & \multicolumn{2}{|c|}{Activity Indicator} & Telescope \\
(BJD)              & (m/s)    & (m/s)   & S-Index & FWHM      &          \\ \hline
2455312.08         & -3.89    & 1.53    &  0.4420 &           & HIRES    \\ \hline
2455373.83         & -3.36    & 1.57    &  0.4130 &           & HIRES    \\ \hline
2456245.38         & -2.86    & 2.84    &         & 7629      & HARPS-N    \\ \hline
2456397.61         & 4.52     & 2.30    &         & 7767      & HARPS-N    \\ \hline

\end{tabular}
\caption{A Few lines of our RV Table are shown here, but the full machine-readable table will be available online.The activity indicator reported for HIRES values is the S-Index while the activity indicator reported for HARPS-N values is the FWHM.}
\label{table:rvs}
\end{center}
\end{table}

\subsection{Simple Keplerian Orbital Fit}

We used the open source Python package \texttt{RadVel} \citep{2018PASP..130d4504F} to model the RVs. We first attempted to measure the mass of Kepler-102d and Kepler-102e by fitting a Keplerian orbit only, in which the RV curve is described by the orbital period, conjunction time, eccentricity, argument of periastron, and RV semi-amplitude of each planet. We include two additional terms to fit the RVs: a zeropoint offset ($\gamma$) and a RV jitter term ($\sigma_{j}$) for each data-set (and two for HARPS-N before and after the CCD change). Jitter accounts for additional Gaussian and non-Gaussian noise that can be astrophysical in origin, or can come from systematics of the spectrograph. This additional uncertainty gets added in quadrature with the intrinsic uncertainties on the RVs to minimize the likelihood function. The likelihood function used in \texttt{RadVel} is:
\begin{equation}
    \ln(\mathcal{L}) =-\sum\limits_{i} \frac{(v_{i}-v_{m}(t_{i}))^{2}}{2(\sigma_{i}^{2}+\sigma_{\rm{jit}}^{2})} - \ln \sqrt{2\pi (\sigma_{i}^{2}+\sigma_{\rm{jit}}^{2})} 
\end{equation}
where $\mathcal{L}$ is the Likelihood, $v_{i}$ and $\sigma_{i}$ are the $i$th radial velocity measurement and its associated uncertainty, $v_{m}(t_{i})$ is the Keplerian model radial velocity at time $t_{i}$, and $\sigma_{\rm{jit}}$ is the jitter estimate.

We fit the RV curve in terms of the two larger planets in the system on Keplerian orbits. We used the published inclinations of the planets, and assumed eccentricities $e=0$, based on findings that the eccentricities in compact, high-multiplicity coplanar systems tend to be significantly lower than what is required for dynamical stability \citep[e.g.,][]{2021AJ....162...55Y}. The orbital periods and conjunction times are precisely known from \textit{Kepler} and were therefore fixed in our analysis \citep{2014ApJS..210...20M, 2019RAA....19...41G}. After optimizing for the maximum likelihood fit, we ran \texttt{RadVel}'s built-in Markov Chain Monte Carlo algorithm \citep{2013PASP..125..306F} to explore the surrounding parameter space and estimate the uncertainty in the model parameters, and to explore the covariance between parameters. 
 
We fit the HIRES and HARPS-N data first separately, and then together for Kepler-102d and Kepler-102e. The semi-amplitudes for each fit, the resulting masses, and the bulk densities of each planet are listed in Table \ref{table:results}. 

While the uncertainties for the mass of Kepler-102d are large, comparing the best-fit values obtained with each dataset is useful to understand the potential effects of stellar variability on the RVs. The HIRES data produces a mass upper limit, defined as the 95th percentile of the posterior distributions, of M$_{d}<$8.3 M$_{\oplus}$, and best-fit mass of M$_{d}$=4.5 $\pm$ 1.9 M$_{\oplus}$. This translates to a best-fit density of $\rho_{d}$=10 $\pm$ 5 g/cm$^{3}$. The HARPS-N data, on the other hand, produces an upper mass limit of M$_{d}<$3.9  M$_{\oplus}$, a best fit mass of M$_{d}$=0.7 $\pm$ 1.6 M$_{\oplus}$ and a density of $\rho_{d}$=2 $\pm$ 4 g/cm$^{3}$. While these best-fit masses are only $\sim$ 1.5 $\sigma$ apart due to large uncertainties, one suggests a composition similar to Mercury, while the other suggests an extensive volatile envelope. Combining the datasets gives a mass upper limit of M$_{d}<$5 M$_{\oplus}$, a best-fit mass of M$_{d}$=2.4 $\pm$ 1.3 M$_{\oplus}$, and a density of $\rho_{d}$=5.4 $\pm$ 3.0 g/cm$^{3}$. This is a density typical of rocky planets--however it is the result of combining the large density measured from HIRES with the low density measured from HARPS-N. We discuss attempts to reconcile the different RV results in Sections 3.3 and 3.4. 

For Kepler-102e, the best-fit values of the semi-amplitude based on each individual dataset, and combined datasets, are listed in Table \ref{table:results}. These are used to calculate a mass of M$_{e}$=5 $\pm$ 2 M$_{\oplus}$ and a density of  $\rho_{e}$=2 $\pm$ 1 g/cm$^{3}$ using HIRES, and a mass of M$_{e}$=3 $\pm$ 2 M$_{\oplus}$ and a density of  $\rho_{e}$=1 $\pm$ 1 g/cm$^{3}$ using HARPS-N. These values differ from each other by 1$\sigma$. 

To test the robustness of our two-planet model, we varied the number of planets included. We use the Bayesian Information Criterion (hereafter BIC) to compare models of this system with two versus five planets. The BIC rewards models that fit the data well (based on optimizing a likelihood function), but also penalizes models based on the number of free parameters. Thus, the BIC can be used to select the simplest model that adequately describes the data. The BIC is defined as:
\begin{equation}
    \text{BIC}=\ln(n)k-2\ln(\mathcal{L}_{\rm{max}})
\end{equation}
where n is the number of data points, $k$ is the number of free parameters, and $\mathcal{L}_{\rm{max}}$ is the maximum value of the likelihood function for the model \citep{doi:10.1080/01621459.1995.10476572}. 

We find that the model with two planets produces a BIC of 822, while the model with five planets produces a BIC of 831. A $\Delta$BIC of 9 indicates the two-planet fit is preferred. This does not suggest that there are only two planets in the system, since we know of 5 transiting planets. It is unsurprising that these three planets are not contributing to the fit, since we would expect their semi-amplitudes to range from 0.03-0.09 m/s assuming masses from the \cite{2014ApJ...783L...6W} mass-radius relation.

\subsection{Gaussian Process RV Fit}
Despite the fact that the combined dataset produces a reasonable mass for Kepler-102d, it prompts the question: why are the best fit semi-amplitudes to the HIRES and HARPS-N datasets so discrepant? One potential reason is stellar activity, which can create time-dependent offsets in the RVs for each dataset. This is not the first time discrepant mass values have been found for exoplanets with HIRES and HARPS-N, another example being Kepler-10 \citep{2014ApJ...789..154D, 2016ApJ...819...83W}. 

To improve our understanding of the system, we applied our knowledge of the stellar activity to construct a physically motivated GP model for the RV variability. We used the posterior values for $\eta_{2}$, $\eta_{3}$, and $\eta_{4}$ from our photometric fit (see section \ref{sec:training}, values listed in Table \ref{table:hyperparams}) as the starting values for a new quasi-periodic kernel, fitting the GP now to the RVs. Because the best-fit values for exponential length $\eta_{2}$ varied from 22 to 37 days in the photometric fit, we fit the RVs with four different starting values for $\eta_{2}$: 22, 27, 31, and 37, which were allowed to vary. The values for the semi-amplitude of each planet remained constant with differing values for $\eta_{2}$. We used 31 days for the fit presented here, but note that any of these values for $\eta_{2}$ would produce the same result. 

In addition to using the photometric fit as starting values, we used Gaussian priors on each value centered on the photometric best-fit  with uncertainties from the photometric fit. To ensure that using the priors for the hyper-parameters on the stellar activity, obtained from the photometry in Section 2.3, are not leading to over-fitting of the RVs, we performed some additional tests with loosened constraints on the activity. Specifically, we removed priors on $\eta_{2}$ and $\eta_{4}$, hence only constraining the rotation period, $\eta_{3}$. When doing this, the masses of Kepler-102d and Kepler-102e remain consistent with those quoted in Table \ref{table:results}. 

\begin{figure}[ht!]
\includegraphics[width=0.5\textwidth]{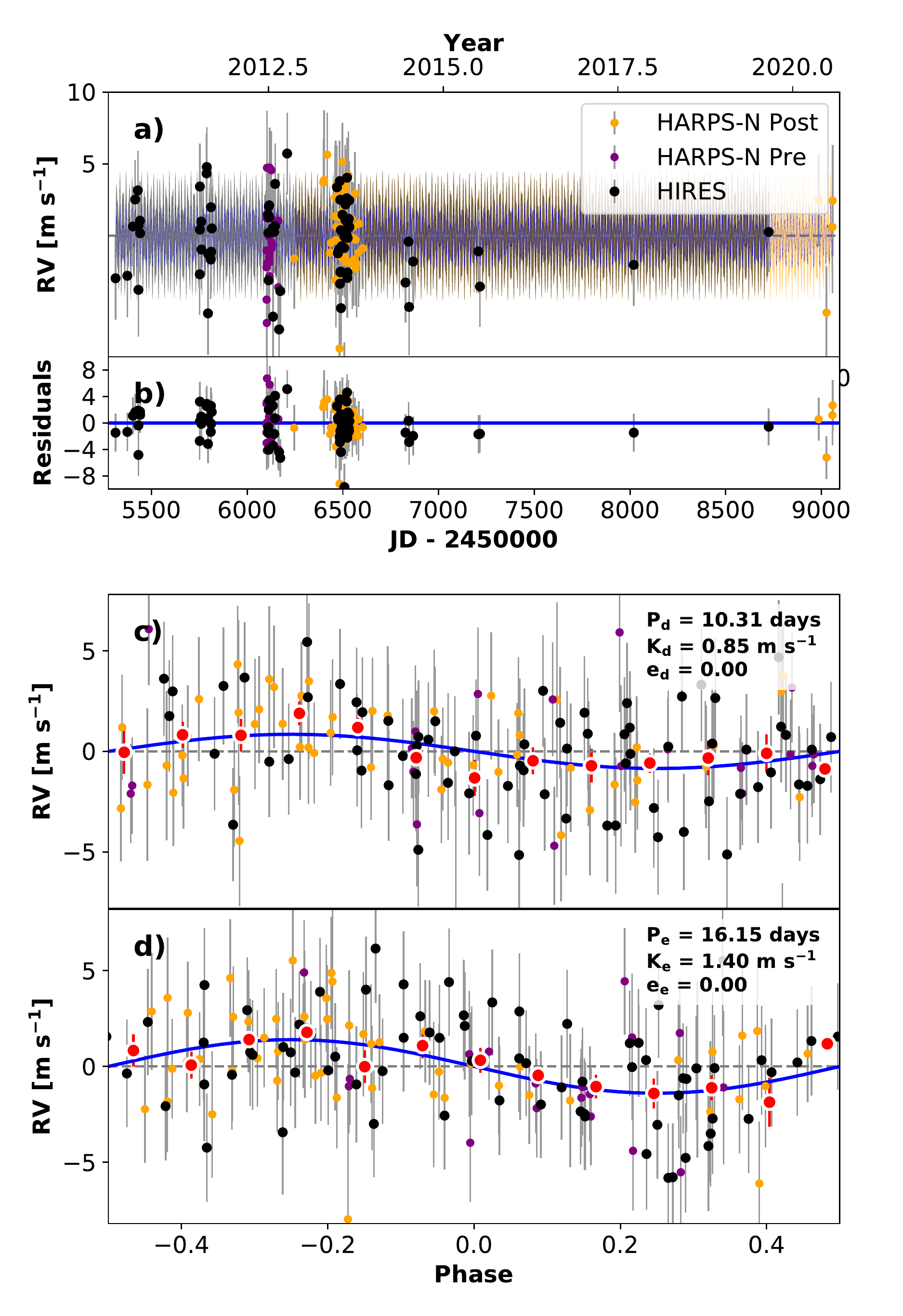}
\caption{Combined Keplerian orbit model for the planets and GP stellar noise fit. The best-fit kernel priors were found using the \textit{Kepler} photometry in Figure \ref{fig:photom} before application to the RVs. a) Radial Velocity vs time as observed by HIRES and HARPS-North (pre and post CCD upgrade) are plotted with 1$\sigma$ uncertainties. The blue line represents the total model of the planets and the Gaussian process regressions, with the grey region showing 1$\sigma$ uncertainties.  b) Residuals of the model subtracted from the RVs c) The phase folded radial velocities for Kepler-102d with the best-fit semi-amplitude model overplotted. The red points are phase-folded weighted mean RVs and their uncertainties. d) same as c) but for Kepler-102e.}
\label{fig:gpfit}
\end{figure}

In addition to the GP stellar noise model, we include a Keplerian RV model for the planets (as described above) and fit for both planetary and stellar signals simultaneously. We included only the two largest planets because our previous BIC analysis demonstrated that the goodness of fit did not benefit from incorporating the three smaller planets. The results of this fit are shown in Figure \ref{fig:gpfit}, and the best-fit values for each planet are given in Table \ref{table:results}.

The best-fit GP hyper-parameters from the posterior of the RV fit are shown in Table \ref{table:hyperparams} together with those from the photometric fit. The parameter with the largest difference between the photometry and radial velocity fits is the variability amplitude. This is expected because the amplitude of the photometry is in units of normalized flux, while the amplitude of the RVs is m/s, and therefore should not be comparable. 

We also used \texttt{PyOrbit} \citep{2016A&A...588A.118M, 2018AJ....155..107M}, utilizing the affine-invarient ensemble sampler \texttt{emcee}, and a DE-MCMC technique \citep{2013PASP..125...83E, 2015A&A...575A..85B} to independently fit the RVs. We recovered values within 1$\sigma$ of the values found with \texttt{RadVel} for the semi-amplitudes of each planet and for the hyper-parameters. The values listed in Tables \ref{table:hyperparams} and \ref{table:results} are those found using \texttt{RadVel}. 

\begin{table}[h]
\begin{center}
 GP Kernel Parameters
\begin{tabular}{|c|c|c|}

\hline
Parameter              & Light curve  Fit     & Radial Velocities Fit\\ \hline
$\eta_{1}$ (Flux, m/s) & 0.0012 $\pm$0.0006   &2.4 $\pm$ 0.4   \\ \hline
$\eta_{2}$ (days)      & 30 $\pm$ 6           & 31 $\pm$ 2   \\ \hline
$\eta_{3}$ (days)      & 26 $\pm$ 1           & 27 $\pm$ 1   \\ \hline
$\eta_{4}$             & 0.2$\pm$0.1          & 0.3 $\pm$ 0.1  \\ \hline
\end{tabular}
\caption{Best fit hyper-parameters for the Quasi-Periodic kernel are shown for both the photometric and radial velocity GP regression.}
\label{table:hyperparams}
\end{center}

\end{table}

\begin{table*}
\begin{center}
    Model Comparison
\begin{tabular}{|c|c|c|c|c|c|c|c|}

\hline
Parameter                        & Previous       & HIRES           & HIRES         & HARPS-N         & HARPS-N         & Full Dataset  & Full Dataset    \\
                                 & Value          &  Without GP     & With GP       & Without GP      & With GP         & Without GP    & With GP         \\  \hline
Kepler-102d K (m/s)              & 1.4$\pm$ 0.6   & 1.5 $\pm$ 0.7   & 0.9 $\pm$0.7  & 0.25 $\pm$ 0.5  & 0.8 $\pm$ 0.5   & 0.8 $\pm$0.4  & 0.9 $\pm$ 0.5 \\ \hline
Kepler-102d Mass (M$_{\oplus}$)  & 3.8$\pm$ 1.8   & 4.5  $\pm$ 1.9  & 2.7$\pm$ 2.1  & 0.7  $\pm$ 1.6  & 2.3 $\pm$ 1.4   & 2.4 $\pm$1.3  & 2.5 $\pm$ 1.4 \\ \hline
Kepler-102d Density (g/cm$^{3}$) & 9.5 $\pm$ 6.5  & 10.1 $\pm$ 4.5  & 6.2 $\pm$ 4.7 & 1.7 $\pm$ 3.9   & 5.6 $\pm$ 3.4   & 5.4 $\pm$3.0  & 5.6 $\pm$ 3.2 \\ \hline
Kepler-102e K (m/s)              & 2.8 $\pm$ 0.6  & 1.4 $\pm$ 0.6   & 1.6 $\pm$0.7  & 0.8  $\pm$ 0.6  & 1.2 $\pm$ 0.7   & 1.2 $\pm$0.4  & 1.4 $\pm$ 0.5 \\ \hline
Kepler-102e Mass (M$_{\oplus}$)  & 8.9$\pm$ 2.0   & 4.8  $\pm$ 2.1  & 5.3 $\pm$2.2  & 2.8 $\pm$ 1.9   & 4.1  $\pm$ 2.2  & 4.0 $\pm$1.4  & 4.7 $\pm$ 1.7 \\ \hline
Kepler-102e Density (g/cm$^{3}$) & 3.5 $\pm$ 1.1  & 1.9 $\pm$ 0.8   & 2.7$\pm$ 0.8  & 1.1 $\pm$ 0.8   & 1.6  $\pm$ 0.8  & 1.6 $\pm$0.6  & 1.8 $\pm$ 0.7 \\ \hline
RV Jitter (m/s)                  &                & 2.4 $\pm$ 0.5   & 1.8 $\pm$ 0.4 & 2.9  $\pm$ 0.8  & 1.9  $\pm$ 1.0  & 3.6 $\pm$ 0.5$^{a}$& 2.5 $\pm$ 0.5$^{a}$      \\  
                                 &                &                 &               &                 &                 & 2.1 $\pm$ 1.0$^{b}$& 2.0 $\pm$ 1.0$^{b}$      \\ \hline 

BIC                              & N/A            & 421             & 418           & 399             & 405             & 822           & 819             \\ \hline

\end{tabular}
\caption{The semi-amplitudes, masses, and densities of Kepler-102d and Kepler-102e across different models and datasets are given above. Previously published masses are taken from \citet{2014ApJS..210...20M}, and densities are calculated using the radii from Table \ref{table:planet}. All models include only the two largest of the five planets in the system. We adopt the values obtained from the Full Dataset with the GP as the semi-amplitudes, masses, and densities for the planets. Jitter key: a=jitter values for the HIRES dataset, b=jitter values for the HARPS-N dataset.}
\label{table:results}
\end{center}

\end{table*}

\begin{figure}[ht!]
\includegraphics[width=0.5\textwidth]{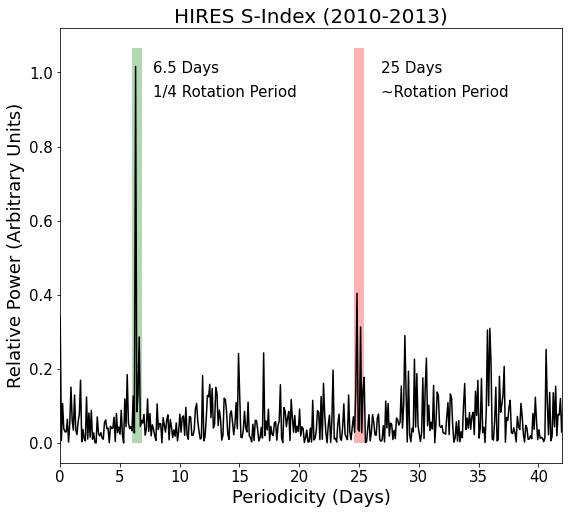}
\includegraphics[width=0.5\textwidth]{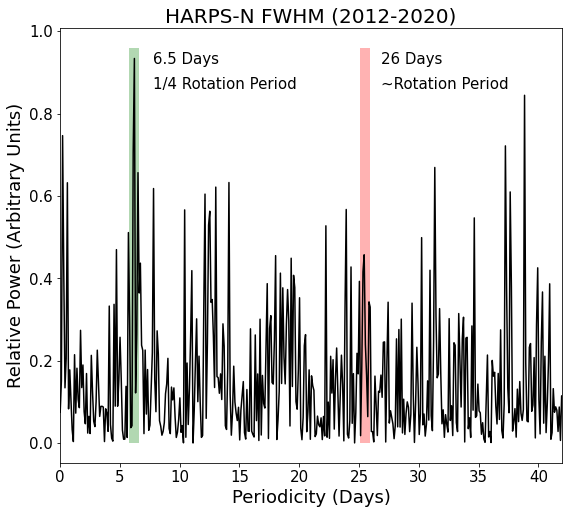}

\caption{The Lomb-Scargle Periodogram of the activity indicators for each RV dataset (S-Index for HIRES, FWHM for HARPS-N) show a notable peak at 6.5 days, which is 1/4 the rotation period (or third harmonic of the rotation period) of the star. In the HIRES S-Index LSP we identify a very tentative signal at 25 days (within 1 standard deviation of the rotation period of the star), while in the HARPS-N FWHM LSP there is not an identifiable signal at the rotation period. The strong peak at the third harmonic of the rotation period suggests that the rotation period of the star is consistent over the timescale of RV data collection and is consistent with the rotation period derived from the Kepler photometry
}
\label{fig:lsp}
\end{figure}

We calculate the BIC of the fit before and after using GP regression to model stellar variability and find that it drops from 822 to 819. A $\Delta$BIC=3, corresponds with positive evidence that the lower BIC model is a better fit relative to the number of free parameters \citep{2019Lorah}, meaning here that the model that includes GP regression for stellar noise is preferred by the BIC even though it includes four additional free parameters. The BIC is not our only metric of determining the validity of the GP: we know from the photometry and the Mt. Wilson S-Value that Kepler-102 shows clear signs of star spots and rotation, and therefore the use of the GP is physically motivated. We also note that lomb-scargle periodograms of the activity indicators for both datasets (S-Index for HIRES, FWHM for HARPS-N) show periodicity at 1/4 the rotation period of the star (6.5 days), and the S-Index from HIRES show periodicity at the full rotation period of the star (25 days) as shown in Figure \ref{fig:lsp}. This suggests that RV variability on the timescale of the rotation period of the star is present in the RVs, and supports the usage of a Quasi-Periodic GP.

\begin{figure*}[ht!]
\includegraphics[width=1.0\textwidth]{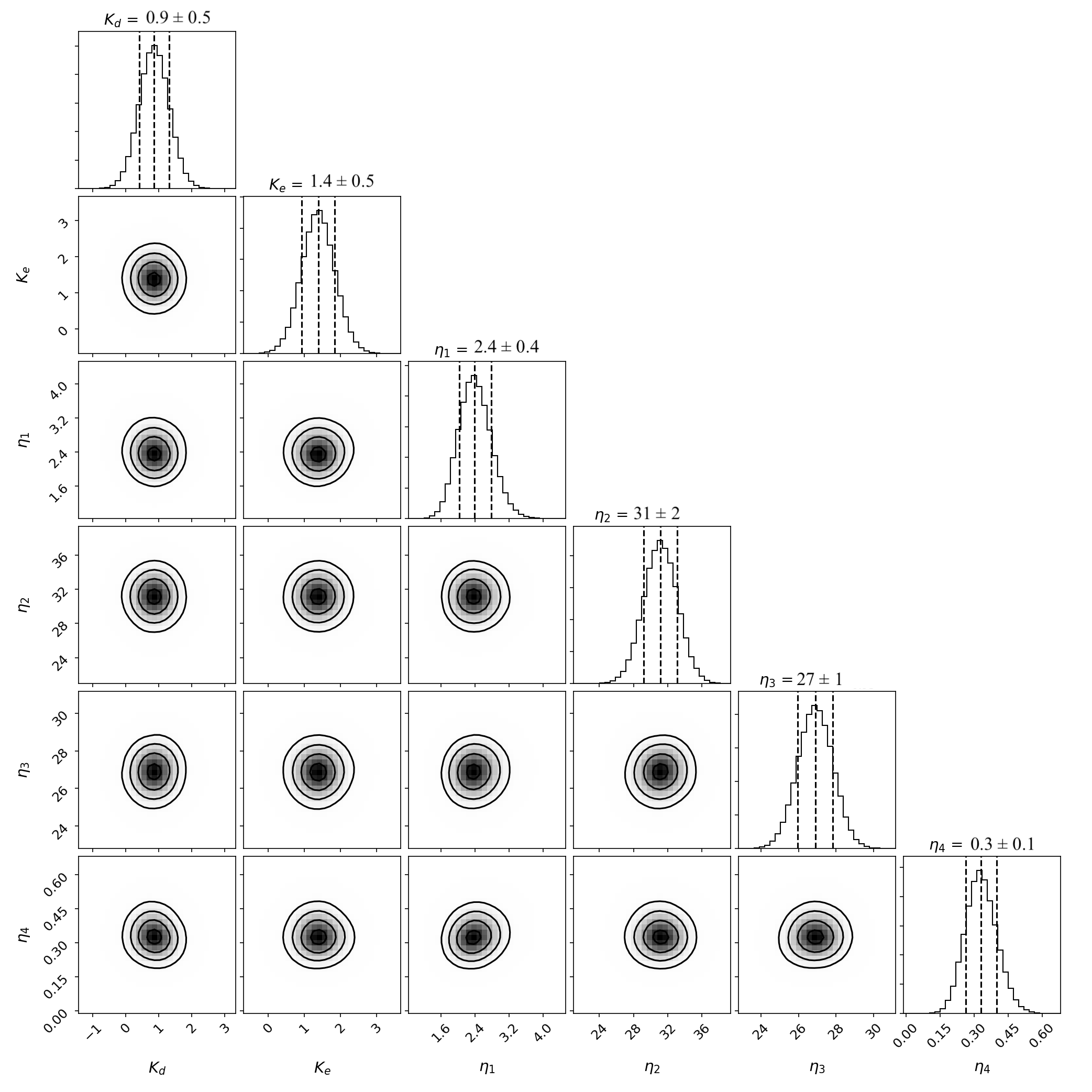}
\caption{Corner plot showing the covariance of the semi-amplitudes of Kepler-102d and Kepler-102e with the four hyper-parameters from the quasi-periodic stellar noise model. The median value of each parameter is shown atop the distribution of values explored through an MCMC plotted as a function of likelihood. The dashed lines through the distributions show the median value as well as the 1$\sigma$ bounds on the parameter. The co-variance plots show the likelihood of models with the parameters varying as a function of one another, and show the 0.5, 1.0, 1.5, and 2.0 $\sigma$ contours. K$_{d}$, K$_{e}$, and $\eta_{1}$ are in units of m/s, while $\eta_{2}$ and $\eta_{3}$ are in units of days.}
\label{fig:cornerplot}
\end{figure*}

A common concern is that the GP can remove the planetary signal. To ascertain whether this was the case in Kepler-102, we examined the covariance between the planet semi-amplitudes and the GP hyperparameters (Figure \ref{fig:cornerplot}). We observe that the semi-amplitudes for the planets are uncorrelated with the GP hyperparameters, indicating that the GP is not artificially reducing nor inflating the semi-amplitude of either planet. This, combined with our exploration of the exponential decay timescale )$\eta_{2}$ discussed above, gives us confidence that the GP is only fitting the stellar noise, and that our measured semi-amplitude reflects the portion of the RV variations due to the orbit of the planets.

\subsection{Discussion}

After accounting for stellar noise, the semi-amplitude for Kepler-102d is only marginally detected, with $\sim$2$\sigma$ significance, with an upper limit on the mass of M$_{d}<$5.3 M$_{\oplus}$ (calculated as the 95th percentile of the posterior distributions). The best-fit semi-amplitude of Kepler-102d is K$_{d}$=0.9 $\pm$ 0.5 m/s. This results in a mass of M$_{d}$=2.5 $\pm$ 1.4 M$_{\oplus}$ compared to the previously published value of 3.8 $\pm$ 1.8 M$_{\oplus}$. Using the planet radius from Table \ref{table:planet}, we get a density of $\rho_{d}$=5.6 $\pm$ 3.2 g/cm$^{3}$--roughly half the previously measured density of 9.5 g/cm$^{3}$. The density measurements from the separate HIRES and HARPS-N datasets produce values within 1 $\sigma$ of each other and the combined dataset using a GP stellar noise model, while they both differ by $>$ 1 $\sigma$ from each other and from the combined dataset without using a GP. Accounting for stellar variability appears to partly reconcile the difference in mass measurements from HIRES and HARPS-N, suggesting that stellar activity could be the source of the discrepant measurements. 

The best-fit semi-amplitude for Kepler-102e in the 2-planet fit with GP regression is K$_{e}$=1.4 $\pm$ 0.5 m/s. The resulting mass for this planet is M$_{e}$=4.7 $\pm$ 1.7 M$_{\oplus}$, lighter than the previously published mass of 8.9 M$_{\oplus}$ \citep{2014ApJS..210...20M}. Using the radius of this planet (Table \ref{table:planet}), we find a density of $\rho_{e}$=1.8 $\pm$ 0.7 g/cm$^{3}$, compared to the 4.7 g/cm$^{3}$ published previously. The density measurements for this planet made using the separate HARPS-N and HIRES datasets, both with and without the GP model, fall within 1 $\sigma$ of this measurement.

\begin{figure*}[ht!]
\includegraphics[width=1\textwidth]{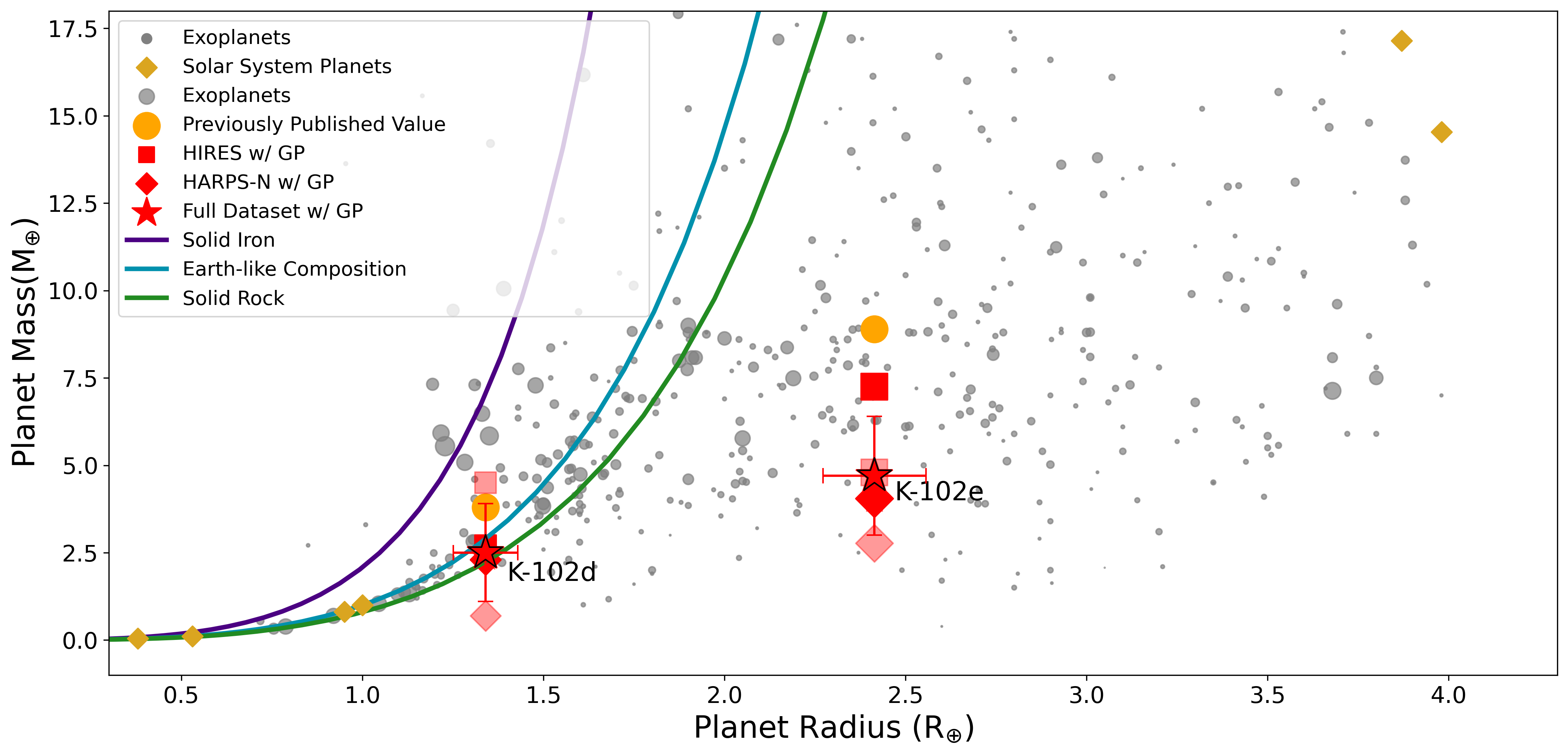}
\includegraphics[width=1\textwidth]{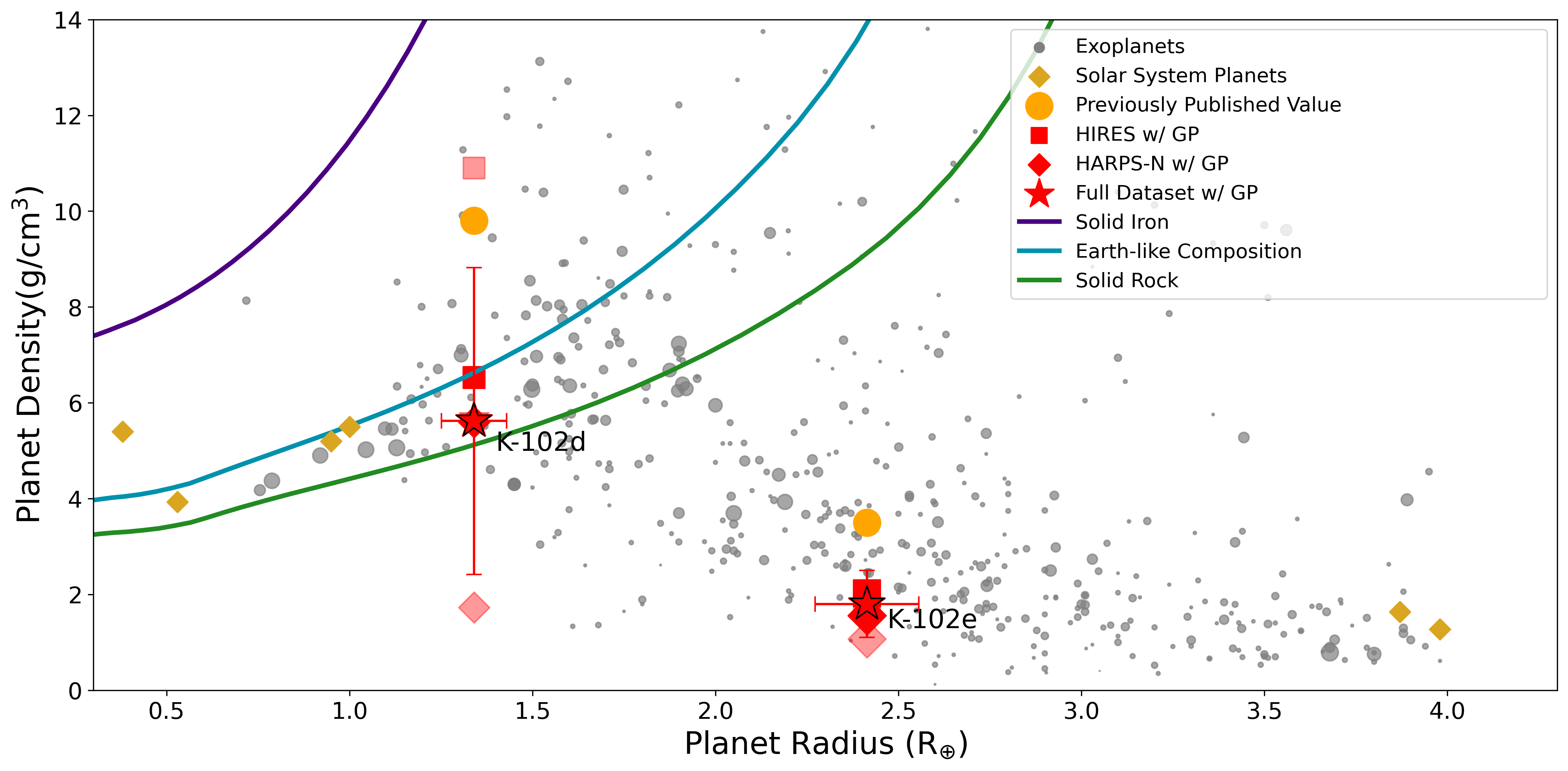}
\caption{Mass (top panel) and Density (bottom panel) are given as a function of radius for all known exoplanets with radius R$<$4R$_{\oplus}$ that have RV mass measurements (shown in grey). The size of the points (excluding Kepler-102d and Kepler-102e) scales inversely with the error of their mass measurement. The black line shows the mass-radius relation from \citet{2014ApJ...783L...6W}. The density-radius curves for planets of solid iron, solid rock, and an Earth-like composition from \cite{2019PNAS..116.9723Z} are shown as well. Kepler-102d and Kepler-102e  (K-102d and K-102e) are shown with their previous mass measurements, and for each model we explore in this paper (Individual vs combined datasets, with and without GP stellar noise modeling). For Kepler-102d, the marker for combined dataset without GP, as well as the marker for HARPS-N with GP, fall behind the marker for full dataset with GP.}
\label{fig:density}
\end{figure*}

Figure \ref{fig:density} places our new mass and density measurements for the Kepler-102 planets in context with the RV-measured densities of other planets. The sample contains all planets with radii R$<$4 R$_{\oplus}$ and published RV mass measurements (NASA Exoplanet Archive \cite{2013PASP..125..989A}, queried 10/01/21). We excluded planets with upper limit mass values only, or mass measurements from Transit Timing Variations. The mass-radius and density-radius trends for rocky and gaseous planets from \cite{2014ApJ...783L...6W} are also included, along with mass-radius and density-radius curves for planets of different compositions from \cite{2019PNAS..116.9723Z}. 

The density of Kepler-102d using the combined dataset and GP stellar noise modeling falls below the trend for average planet density at its given radius, while the uncertainty on this measurement places it within 1$\sigma$ of this relation. The density of Kepler-102e using the full dataset and GP modeling falls below the average density relation for sub-Neptunes, but is also consistent with it to within 1$\sigma$. 

The wide variety of densities found using different datasets and models illustrates the difficulty of measuring precise densities for small planets. However, the individual datasets used without GPs fall the furthest away from the expected density range for rocky planets, while those same datasets used with GPs fall much closer to that range and fall closer to each other and the combined dataset. This suggests that the GP analysis helps with constraining more accurate densities for rocky planets, and should be implemented with care for systems with clear signals of rotation and activity in the host star. 

\section{Planet Compositions}
\label{sec:composition}

\subsection{Planet Interior Modeling}

We combined our mass and radius measurements of Kepler-102d and 102e with models of the interior structures of differentiated rocky planets to constrain their compositions. We assumed the planets are differentiated, a consequence of energy released and melting occurring during accretion \citep{Chao2021}, but the mass-radius relation of rocky planets is not very sensitive to their degree of differentiation \citep{2008ApJ...688..628E}.  We also ignored the contribution of any atmosphere to the radius, but this could be significant in the case of thick hydrogen-helium envelopes. By ignoring such atmospheres we estimated \emph{maximum} model masses; a model mass which significantly exceeds the measured value for the lightest plausible composition indicates that such an atmosphere is present.  

Our modeling used {\tt BurnMan 0.9} \citep{2016ascl.soft10010C}, which takes user-provided equations of state (EOS) and the masses of individual layers in a differentiated planet and computes the inner and outer radii of each layer. We incorporated {\tt BurnMan} in an iterative scheme which estimates the mass of a planet with a given (relatively precisely known) radius, and specified mass distributions between the different layers (in this case, a solid metallic inner core, a liquid metallic outer core, a silicate mantle, and an outer mantle of high-pressure H$_2$O ice phases). The iteration begins with an educated guess for the outer radius of each layer based on a power-law scaling of the Preliminary Earth Reference Model \citep{1981PEPI...25..297D} scaling with the specified mass fractions. The mass of each layer is computed based on the outer radius of that layer and the outer radius of the underlying layer (the radius of the outermost layer is set to the known transit value), the mass fractions of each layer are then re-computed, the radii of the individual layers are adjusted based on these mass fractions, and the iteration continues until the {\tt BurnMan}-calculated mass fractions fall within a given tolerance (0.1\%) of the specified values.  After convergence, the planet mass--which is the sum of the masses of the individual layers--is compared to measured values.  

The silicate mantle is modeled with the EOS of Mg-bridgmanite \citep{2011GeoJI.184.1180S}, the inner core with the EOS of solid Fe  \citep{2006PhRvL..97u5504D}, and the outer core with the EOS of liquid Fe  \citep{1994JGR....99.4273A}. The ice mantle is modeled entirely as Ice VII; layers of liquid H$_2$O and Ice Ih could exist near the surface, but would not contribute significantly ($<$100 km) to the planet radius. For the Ice VII EOS, a third-order Birch-Murnaghan formulation was implemented \citep{FRANK20042781}. Ice X could exist at pressures exceeding 60 GPa \citep{BrownJM2020}, but the EOS of this phase is poorly constrained and so we use that of Ice VII in its place.  In this way, the mass of a planet with a known radius can be predicted as a function of the mass fraction of a core, silicate mantle, and ice mantle. 

\begin{figure}
  \includegraphics[width=0.5\textwidth]{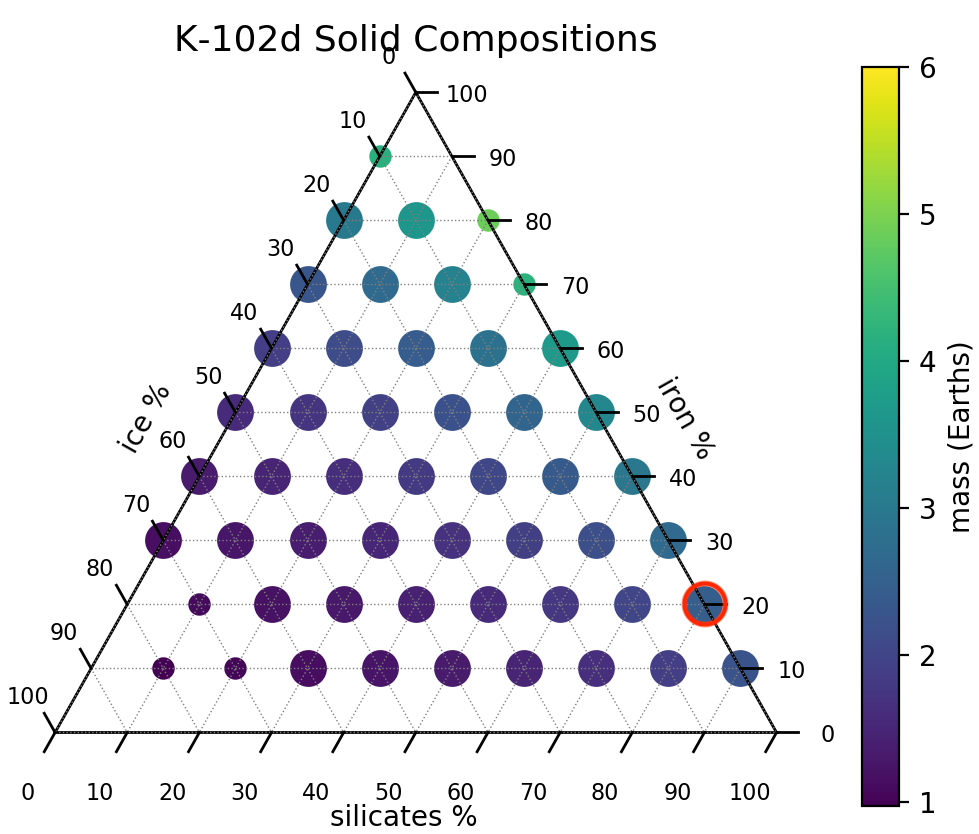}  \\
\caption{Ternary plot for Kepler-102d (K-102d) showing the range of possible combinations of iron, rock, and ice in increments of 10$\%$ for a planet of fixed radius. The colorbar shows the resulting mass of such a planet. Masses and volumes of different substances are calculated by solving equations of state using assumed internal pressure and temperatures using \texttt{Burnman} 0.9 \citep{2016ascl.soft10010C}. Every point corresponds to a planet with physically allowed equations of state for that combination of iron, rock, and ice--and all produce values for the mass of the planet within 1$\sigma$ (large points) and 2$\sigma$ (small points) of our measured mass value for that planet. The horizontal dotted lines show iron fraction, the dotted lines with negative slope show ice fraction, and the lines with positive slope show silicate fraction. The red circle highlights the composition that produces our best-fit mass of M$_{d}$=2.5M$_{\oplus}$. }
\label{fig:composition}
\end{figure}

Figure \ref{fig:composition} shows the range of possible compositions for Kepler-102d as a ternary diagram. The color bar shows the mass for a specified composition (as a combination of iron, silicate, and ice fractions) at the radius of the planet (from Table \ref{table:planet}). We recovered a wide range of possible compositions for a planet of radius of 1.34 R$_{\oplus}$, and with a mass within 2$\sigma$ of our best-fit value. These are also all possible compositions consistent with the mass upper limit of M$_{d}<$5.3 M$_{\oplus}$. While the vast majority of ice, iron, and rock fractions produce planets with a mass within 2$\sigma$ of our measured value, the composition that most closely reproduces a mass of 2.5 M$_{\oplus}$ is a planet of roughly 20$\%$ iron, 80$\%$ rock, and 0$\%$ ice by mass. This is a slightly smaller core mass fraction than the Earth ($\sim$30$\%$ iron), but consistent with the Earth within 1$\sigma$. It is important to note, however, that the composition is highly unconstrained due to the large uncertainty on the mass measurement. 

For Kepler-102e, we were unable to identify a combination only of iron, rock, and ice that produces a mass within 3$\sigma$ of the best-fit value. Therefore, this planet must have a non-solid component as most sub-Neptune sized planets do, likely a gaseous envelope. 

\subsection{Core-Envelope Model for Kepler-102e}
Kepler-102e has a density consistent with other sub-Neptunes, and therefore a significant portion of the planet's radius likely comes from this atmosphere rather than a solid surface \citep{2015ApJ...801...41R}. However, a planet with this mass and radius can be estimated as having only $\sim$2$\%$ of its mass in the H/He envelope using the relation between radius and envelope fraction from \cite{2014ApJ...792....1L}. 

As a first order approximation, we could assume that the core of the planet (the solid interior portion of the planet inclusive of rock and iron) follows the mass-radius relation for terrestrial planets: 
\begin{equation}
    R_{\rm core}\approx (\frac{M_{\rm core}}{M_{\oplus}})^{0.25}\approx (\frac{M_{P}}{M_{\oplus}})^{0.25}
\end{equation}
\\

where $M_{core}$ is the mass of the planet's solid core, and $M_{P}$ is the mass of the entire planet \citep{2014ApJ...792....1L}. This relation assumes an Earth-like composition without significant quantities of water or ice. It has an inherent uncertainty of 10$\%$ when rock/iron fractions are allowed to vary. 

\begin{figure}[ht!]
\includegraphics[width=0.5\textwidth]{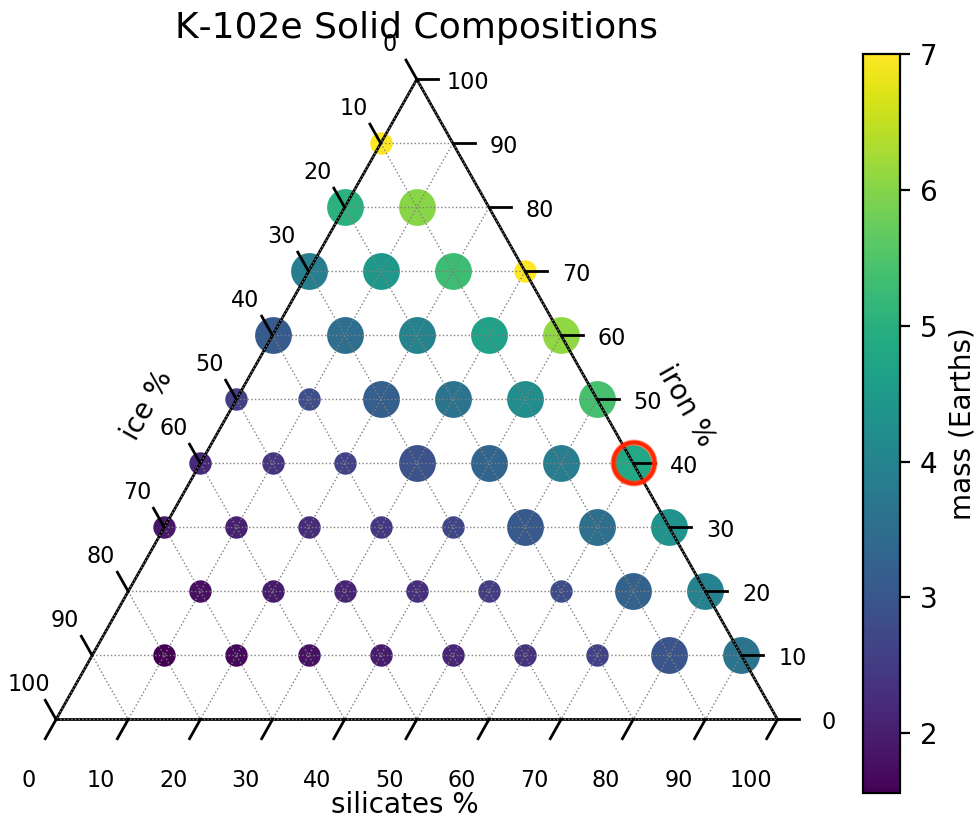}
\caption{Ternary plot for Kepler-102e (K-102e) solid core compositions, assuming a core radius of R$_{c}$=1.5 R$_{\oplus}$. Large points represent planet compositions consistent with our measured mass of the planet to within 1$\sigma$, while smaller points represent compositions within 2$\sigma$. The planet core composition that most closely matches our measured value of M$_{e}$=4.7 M$_{\oplus}$ is one of 40$\%$ iron and 60$\%$ silicate rock--more dense than the best-fit composition for Kepler-102d.}
\label{fig:coreradius}
\end{figure}

Using this method, we can approximate the radius of the core of Kepler-102e as R$_{c}$=1.5 R$_{\oplus}$, corresponding to a density of $\rho_{c}$=9 g/cm$^{3}$. The results of building a core of this radius and a mass within 2$\sigma$ of our best-fit value are shown in Figure \ref{fig:coreradius}, accounting for a 2$\%$ reduction in mass due to the envelope. Compositions within 1$\sigma$ of this mass range could have silicate fractions of 0.0-0.9, iron fractions of 0.1-0.6, and ice fractions of 0.2-0.4. 

This model of the core is highly uncertain, however, and does not consider significant ice fractions. Another plausible core composition is one of roughly 50$\%$ water ice, 50$\%$ rock based on Solar System abundances \citep{2003ApJ...591.1220L}.

While we cannot determine the core size, we can set upper and lower limits using our {\tt Burnman} models. Ignoring significant ice fractions in the core \citep{2017ApJ...847...29O, 2015ApJ...801...41R}, the largest core possible for Kepler-102e is one that lacks any iron and contains silicate rock only at R$_{c}$=1.6 R$_{\oplus}$. Considering compositions that include ice, the largest core possible for Kepler-102e is one that contains 80$\%$ ice, 10$\%$ silicate rock, and 10$\%$ iron at R$_{c}$=2.0 R$_{\oplus}$. At the other extreme, the core of Kepler-102e could be as small as R$_{c}$=1.2 R$_{\oplus}$ if made entirely of compressed iron. This puts limits on the extent of the planet's atmosphere--ranging from to 16-50$\%$ of the planet's radius if we consider ice fractions, and 33-50$\%$ if we exclude ice. We can then estimate that the mass of the H/He envelope then falls between 1.8-4$\%$ of the total planet mass \citep{2014ApJ...792....1L}, consistent with the assumption that the envelope mass is very small compared with the total planet mass. 

\subsection{Water and Ice Fractions}
\label{sec:EQT}
To investigate the plausibility of solid, differentiated planets containing significant fractions of water ice at their orbital periods, we can compute the equilibrium temperature of the planets. Assuming that a planet is radiating as a blackbody, we can calculate the equilibrium temperature of the planet as follows:
\begin{equation}
    T_{Planet}=T_{*}[f(1-A_{B})]^{1/4}\sqrt{\frac{R_{*}}{a}}
\end{equation}
\\

Here, $A_{B}$ is the Bond albedo, $f$ is the re-radiation factor, and $a$ is the semi-major axis of the planet. The only unconstrained parameters are the re-radiation factor and the albedo. Albedo ranges from 0.12 to 0.75 for Solar System planets, while $f$ is thought to vary from 1/4-2/3 \citep{2007ApJ...667L.191L}. 

We use the stellar properties from Table \ref{table:stellar} and planet properties from Table \ref{table:planet} to calculate equilibrium temperature. Assuming they both have similar albedos to Earth ($\approx$0.30), or even Neptune ($\approx$0.29), it would give Kepler-102d and Kepler-102e equilibrium temperatures of $\approx$802 K and $\approx$ 693 K, respectively. These are well above the melting (273 K) and boiling (373 K) temperatures of water at standard temperature and pressure (STP), and the temperature of both planets could easily be above the critical point of water (647 K) \citep{2002JPCRD..31..387W}. Water-rich planets that orbit close to their host star are likely to experience a runaway greenhouse effect and photo-dissociation of water molecules \citep{2015AsBio..15..119L, 2017MNRAS.464.3728B}, raising the question of whether they could have retained significant water. 

Unless Kepler-102e has an albedo higher than Enceladus, it would need a significant amount of atmospheric pressure to sustain liquid water or ice on the surface of the planet, on the order of 10 GPa surface pressure \citep{2019arXiv190709598J}. While we are unable to say whether or not Kepler-102e has surface oceans, compressed ice, or even water in the atmosphere, we can say this sub-Neptune sized planet requires a substantial gaseous envelope, and the non-solid portion of the planet cannot be explained by water alone.

\section{Conclusion}
\label{sec:conclusion}
We have applied Gaussian Processes to model the correlated stellar noise of Kepler-102 in both photometry from \textit{Kepler} (Quarters 1-17) and 146 RVs collected using HIRES and HARPS-N. This allowed us to more accurately measure the masses of two of the five transiting planets. We have determined: 
\begin{itemize}
 
    \item Kepler-102d, a super-Earth size planet with a radius of R$_{d}$=1.3 $\pm$ 0.09 R$_{\oplus}$, has a mass upper limit of M$_{d} < $5.3 M$_{\oplus}$ [95\% confidence].  Our best fit to the RVs yields a mass of M$_{d}$=2.5 $\pm$ 1.4 M$_{\oplus}$ and a density of $\rho_{d}$=5.6 $\pm$ 3.2 g/cm$^{3}$. This density is roughly half the value previously determined with the radius from \cite{2018AJ....156..264F} and mass from \cite{2014ApJS..210...20M}. It also falls within 1$\sigma$ of the mass-density relations published by \cite{2020A&A...634A..43O} and \cite{2014ApJ...783L...6W}. The best-fit composition is $\sim$20$\%$ iron $\sim$80$\%$ silicate rock, and is consistent with an Earth-like composition, but because of the low precision achieved in the mass measurement the composition remains highly unconstrained. 
    
    \item Kepler-102e, a sub-Neptune size planet with a radius of 2.41 $\pm$ 0.14 R$_{\oplus}$, has a mass of 4.7 $\pm$ 1.7 and a density of 1.8 $\pm$ 0.7 g/cm$^{3}$. This density is also $\sim$50$\%$ smaller than the previously published value, and is consistent with other observed sub-Neptune densities. Kepler-102e has a hydrogen/helium envelope likely containing $\sim$2-4$\%$ of the mass of the planet. Without significant fractions of water ice, this atmosphere accounts for at least 33$\%$ of the radius of the planet. It has a minimum core size of R$_{c}>$1.2 R$_{\oplus}$, therefore giving the envelope a maximum size of 50$\%$ of the planet radius. 
\end{itemize}
Characterizing activity in stars with clear rotation signals yields more robust planet masses, enabling a more realistic interpretation of planet interiors. Our study has shown that modeling stellar activity using GPs is important to derive masses and densities of small planets, particularly when there are discrepant values measured by different instruments. Using an informed estimate of the rotation period and decay timescale from the photometry is critical to ensure that the GP fit is physically motivated. GP analysis of this sort will be particularly important for more precise RV instruments such as Maroon-X, NEID, ESPRESSO/VLT, KPF, and future instruments for which stellar noise will become even more dominant over instrumental noise. 

\section{Acknowledgements}

This material is based upon work supported by the National Science Foundation Graduate Research Fellowship under Grant No. 1842402. 

 C.L.B., L.W. and D.H. acknowledge support from National Aeronautics and Space Administration (80NSSC19K0597) issued through the Astrophysics Data Analysis Program.

D.H. also acknowledges support from the Alfred P. Sloan Foundation.

KR acknowledges support from the UK STFC via grant ST/V000594/1.

EG acknowledges support from NASA Award 80NSSC20K0957 (Exoplanets Research Program).

The HARPS-N project was funded by the Prodex Program of the Swiss Space Office (SSO), the Harvard- University Origin of Life Initiative (HUOLI), the Scottish Universities Physics Alliance (SUPA), the University of Geneva, the Smithsonian Astrophysical Observatory (SAO), the Italian National Astrophysical Institute (INAF), University of St. Andrews, Queen’s University Belfast, and University of Edinburgh. This work hase been supported by the National Aeronautics and Space Administration under grant No. NNX17AB59G, issued through the Exoplanets Research Program.

Some of the data presented in this paper were obtained from the Mikulski Archive for Space Telescopes (MAST) at the Space Telescope Science Institute. The specific observations analyzed can be accessed via \dataset[10.17909/T9059R]{https://doi.org/10.17909/T9059R}.

 Partly based on observations made with the Italian {\it Telescopio Nazionale
Galileo} (TNG) operated by the {\it Fundaci\'on Galileo Galilei} (FGG) of the
{\it Istituto Nazionale di Astrofisica} (INAF) at the
{\it  Observatorio del Roque de los Muchachos} (La Palma, Canary Islands, Spain).

\software{Lightcurve \citep{2018ascl.soft12013L}, kiauhoku \citep{2020ascl.soft11027C}, RadVel \citep{2018PASP..130d4504F}, PyORBIT \citep{2016A&A...588A.118M, 2018AJ....155..107M}, george \citep{2015ITPAM..38..252A}, emcee \citep{2013PASP..125..306F}, BurnMan 0.9 \citep{2016ascl.soft10010C}, NumPy \citep{harris2020array}, Matplotlib \citep{Hunter:2007}, pandas \citep{mckinney-proc-scipy-2010}, Astropy \citep{astropy:2013, astropy:2018, astropy:2022}, SciPy \citep{2020SciPy-NMeth}}


\bibliography{k102.bib,more_refs.bib}

\begin{thebibliography}{}
\expandafter\ifx\csname natexlab\endcsname\relax\def\natexlab#1{#1}\fi
\providecommand{\url}[1]{\href{#1}{#1}}
\providecommand{\dodoi}[1]{doi:~\href{http://doi.org/#1}{\nolinkurl{#1}}}
\providecommand{\doeprint}[1]{\href{http://ascl.net/#1}{\nolinkurl{http://ascl.net/#1}}}
\providecommand{\doarXiv}[1]{\href{https://arxiv.org/abs/#1}{\nolinkurl{https://arxiv.org/abs/#1}}}

\bibitem[{{Akeson} {et~al.}(2013){Akeson}, {Chen}, {Ciardi}, {Crane}, {Good},
  {Harbut}, {Jackson}, {Kane}, {Laity}, {Leifer}, {Lynn}, {McElroy}, {Papin},
  {Plavchan}, {Ram{\'\i}rez}, {Rey}, {von Braun}, {Wittman}, {Abajian}, {Ali},
  {Beichman}, {Beekley}, {Berriman}, {Berukoff}, {Bryden}, {Chan}, {Groom},
  {Lau}, {Payne}, {Regelson}, {Saucedo}, {Schmitz}, {Stauffer}, {Wyatt}, \&
  {Zhang}}]{2013PASP..125..989A}
{Akeson}, R.~L., {Chen}, X., {Ciardi}, D., {et~al.} 2013, \pasp, 125, 989,
  \dodoi{10.1086/672273}

\bibitem[{{Ambikasaran} {et~al.}(2015){Ambikasaran}, {Foreman-Mackey},
  {Greengard}, {Hogg}, \& {O'Neil}}]{2015ITPAM..38..252A}
{Ambikasaran}, S., {Foreman-Mackey}, D., {Greengard}, L., {Hogg}, D.~W., \&
  {O'Neil}, M. 2015, IEEE Transactions on Pattern Analysis and Machine
  Intelligence, 38, 252, \dodoi{10.1109/TPAMI.2015.2448083}

\bibitem[{{Anderson} \& {Ahrens}(1994)}]{1994JGR....99.4273A}
{Anderson}, W.~W., \& {Ahrens}, T.~J. 1994, \jgr, 99, 4273,
  \dodoi{10.1029/93JB03158}

\bibitem[{{Astropy Collaboration} {et~al.}(2013){Astropy Collaboration},
  {Robitaille}, {Tollerud}, {Greenfield}, {Droettboom}, {Bray}, {Aldcroft},
  {Davis}, {Ginsburg}, {Price-Whelan}, {Kerzendorf}, {Conley}, {Crighton},
  {Barbary}, {Muna}, {Ferguson}, {Grollier}, {Parikh}, {Nair}, {Unther},
  {Deil}, {Woillez}, {Conseil}, {Kramer}, {Turner}, {Singer}, {Fox}, {Weaver},
  {Zabalza}, {Edwards}, {Azalee Bostroem}, {Burke}, {Casey}, {Crawford},
  {Dencheva}, {Ely}, {Jenness}, {Labrie}, {Lim}, {Pierfederici}, {Pontzen},
  {Ptak}, {Refsdal}, {Servillat}, \& {Streicher}}]{astropy:2013}
{Astropy Collaboration}, {Robitaille}, T.~P., {Tollerud}, E.~J., {et~al.} 2013,
  \aap, 558, A33, \dodoi{10.1051/0004-6361/201322068}

\bibitem[{{Astropy Collaboration} {et~al.}(2018){Astropy Collaboration},
  {Price-Whelan}, {Sip{\H{o}}cz}, {G{\"u}nther}, {Lim}, {Crawford}, {Conseil},
  {Shupe}, {Craig}, {Dencheva}, {Ginsburg}, {Vand erPlas}, {Bradley},
  {P{\'e}rez-Su{\'a}rez}, {de Val-Borro}, {Aldcroft}, {Cruz}, {Robitaille},
  {Tollerud}, {Ardelean}, {Babej}, {Bach}, {Bachetti}, {Bakanov}, {Bamford},
  {Barentsen}, {Barmby}, {Baumbach}, {Berry}, {Biscani}, {Boquien}, {Bostroem},
  {Bouma}, {Brammer}, {Bray}, {Breytenbach}, {Buddelmeijer}, {Burke},
  {Calderone}, {Cano Rodr{\'\i}guez}, {Cara}, {Cardoso}, {Cheedella}, {Copin},
  {Corrales}, {Crichton}, {D'Avella}, {Deil}, {Depagne}, {Dietrich}, {Donath},
  {Droettboom}, {Earl}, {Erben}, {Fabbro}, {Ferreira}, {Finethy}, {Fox},
  {Garrison}, {Gibbons}, {Goldstein}, {Gommers}, {Greco}, {Greenfield},
  {Groener}, {Grollier}, {Hagen}, {Hirst}, {Homeier}, {Horton}, {Hosseinzadeh},
  {Hu}, {Hunkeler}, {Ivezi{\'c}}, {Jain}, {Jenness}, {Kanarek}, {Kendrew},
  {Kern}, {Kerzendorf}, {Khvalko}, {King}, {Kirkby}, {Kulkarni}, {Kumar},
  {Lee}, {Lenz}, {Littlefair}, {Ma}, {Macleod}, {Mastropietro}, {McCully},
  {Montagnac}, {Morris}, {Mueller}, {Mumford}, {Muna}, {Murphy}, {Nelson},
  {Nguyen}, {Ninan}, {N{\"o}the}, {Ogaz}, {Oh}, {Parejko}, {Parley}, {Pascual},
  {Patil}, {Patil}, {Plunkett}, {Prochaska}, {Rastogi}, {Reddy Janga},
  {Sabater}, {Sakurikar}, {Seifert}, {Sherbert}, {Sherwood-Taylor}, {Shih},
  {Sick}, {Silbiger}, {Singanamalla}, {Singer}, {Sladen}, {Sooley},
  {Sornarajah}, {Streicher}, {Teuben}, {Thomas}, {Tremblay}, {Turner},
  {Terr{\'o}n}, {van Kerkwijk}, {de la Vega}, {Watkins}, {Weaver}, {Whitmore},
  {Woillez}, {Zabalza}, \& {Astropy Contributors}}]{astropy:2018}
{Astropy Collaboration}, {Price-Whelan}, A.~M., {Sip{\H{o}}cz}, B.~M., {et~al.}
  2018, \aj, 156, 123, \dodoi{10.3847/1538-3881/aabc4f}

\bibitem[{{Astropy Collaboration} {et~al.}(2022){Astropy Collaboration},
  {Price-Whelan}, {Lim}, {Earl}, {Starkman}, {Bradley}, {Shupe}, {Patil},
  {Corrales}, {Brasseur}, {N{"o}the}, {Donath}, {Tollerud}, {Morris},
  {Ginsburg}, {Vaher}, {Weaver}, {Tocknell}, {Jamieson}, {van Kerkwijk},
  {Robitaille}, {Merry}, {Bachetti}, {G{"u}nther}, {Aldcroft},
  {Alvarado-Montes}, {Archibald}, {B{'o}di}, {Bapat}, {Barentsen}, {Baz{'a}n},
  {Biswas}, {Boquien}, {Burke}, {Cara}, {Cara}, {Conroy}, {Conseil}, {Craig},
  {Cross}, {Cruz}, {D'Eugenio}, {Dencheva}, {Devillepoix}, {Dietrich},
  {Eigenbrot}, {Erben}, {Ferreira}, {Foreman-Mackey}, {Fox}, {Freij}, {Garg},
  {Geda}, {Glattly}, {Gondhalekar}, {Gordon}, {Grant}, {Greenfield}, {Groener},
  {Guest}, {Gurovich}, {Handberg}, {Hart}, {Hatfield-Dodds}, {Homeier},
  {Hosseinzadeh}, {Jenness}, {Jones}, {Joseph}, {Kalmbach}, {Karamehmetoglu},
  {Ka{l}uszy{'n}ski}, {Kelley}, {Kern}, {Kerzendorf}, {Koch}, {Kulumani},
  {Lee}, {Ly}, {Ma}, {MacBride}, {Maljaars}, {Muna}, {Murphy}, {Norman},
  {O'Steen}, {Oman}, {Pacifici}, {Pascual}, {Pascual-Granado}, {Patil},
  {Perren}, {Pickering}, {Rastogi}, {Roulston}, {Ryan}, {Rykoff}, {Sabater},
  {Sakurikar}, {Salgado}, {Sanghi}, {Saunders}, {Savchenko}, {Schwardt},
  {Seifert-Eckert}, {Shih}, {Jain}, {Shukla}, {Sick}, {Simpson},
  {Singanamalla}, {Singer}, {Singhal}, {Sinha}, {Sip{H{o}}cz}, {Spitler},
  {Stansby}, {Streicher}, {{{S}}umak}, {Swinbank}, {Taranu}, {Tewary},
  {Tremblay}, {Val-Borro}, {Van Kooten}, {Vasovi{'c}}, {Verma}, {de Miranda
  Cardoso}, {Williams}, {Wilson}, {Winkel}, {Wood-Vasey}, {Xue}, {Yoachim},
  {Zhang}, {Zonca}, \& {Astropy Project Contributors}}]{astropy:2022}
{Astropy Collaboration}, {Price-Whelan}, A.~M., {Lim}, P.~L., {et~al.} 2022,
  apj, 935, 167, \dodoi{10.3847/1538-4357/ac7c74}

\bibitem[{{Basri} \& {Nguyen}(2018)}]{2018ApJ...863..190B}
{Basri}, G., \& {Nguyen}, H.~T. 2018, \apj, 863, 190,
  \dodoi{10.3847/1538-4357/aad3b6}

\bibitem[{{Berger} {et~al.}(2018){Berger}, {Huber}, {Gaidos}, \& {van
  Saders}}]{2018ApJ...866...99B}
{Berger}, T.~A., {Huber}, D., {Gaidos}, E., \& {van Saders}, J.~L. 2018, \apj,
  866, 99, \dodoi{10.3847/1538-4357/aada83}

\bibitem[{{Bolmont} {et~al.}(2017){Bolmont}, {Selsis}, {Owen}, {Ribas},
  {Raymond}, {Leconte}, \& {Gillon}}]{2017MNRAS.464.3728B}
{Bolmont}, E., {Selsis}, F., {Owen}, J.~E., {et~al.} 2017, \mnras, 464, 3728,
  \dodoi{10.1093/mnras/stw2578}

\bibitem[{{Bonomo} {et~al.}(2015){Bonomo}, {Sozzetti}, {Santerne}, {Deleuil},
  {Almenara}, {Bruno}, {D{\'\i}az}, {H{\'e}brard}, \&
  {Moutou}}]{2015A&A...575A..85B}
{Bonomo}, A.~S., {Sozzetti}, A., {Santerne}, A., {et~al.} 2015, \aap, 575, A85,
  \dodoi{10.1051/0004-6361/201323042}

\bibitem[{{Bonomo} {et~al.}(2019){Bonomo}, {Zeng}, {Damasso}, {Leinhardt},
  {Justesen}, {Lopez}, {Lund}, {Malavolta}, {Silva Aguirre}, {Buchhave},
  {Corsaro}, {Denman}, {Lopez-Morales}, {Mills}, {Mortier}, {Rice}, {Sozzetti},
  {Vanderburg}, {Affer}, {Arentoft}, {Benbakoura}, {Bouchy},
  {Christensen-Dalsgaard}, {Collier Cameron}, {Cosentino}, {Dressing},
  {Dumusque}, {Figueira}, {Fiorenzano}, {Garc{\'\i}a}, {Handberg},
  {Harutyunyan}, {Johnson}, {Kjeldsen}, {Latham}, {Lovis}, {Lundkvist},
  {Mathur}, {Mayor}, {Micela}, {Molinari}, {Motalebi}, {Nascimbeni}, {Nava},
  {Pepe}, {Phillips}, {Piotto}, {Poretti}, {Sasselov}, {S{\'e}gransan}, {Udry},
  \& {Watson}}]{2019NatAs...3..416B}
{Bonomo}, A.~S., {Zeng}, L., {Damasso}, M., {et~al.} 2019, Nature Astronomy, 3,
  416, \dodoi{10.1038/s41550-018-0684-9}

\bibitem[{{Borucki} {et~al.}(2010){Borucki}, {Koch}, {Basri}, {Batalha},
  {Brown}, {Caldwell}, {Caldwell}, {Christensen-Dalsgaard}, {Cochran},
  {DeVore}, {Dunham}, {Dupree}, {Gautier}, {Geary}, {Gilliland}, {Gould},
  {Howell}, {Jenkins}, {Kondo}, {Latham}, {Marcy}, {Meibom}, {Kjeldsen},
  {Lissauer}, {Monet}, {Morrison}, {Sasselov}, {Tarter}, {Boss}, {Brownlee},
  {Owen}, {Buzasi}, {Charbonneau}, {Doyle}, {Fortney}, {Ford}, {Holman},
  {Seager}, {Steffen}, {Welsh}, {Rowe}, {Anderson}, {Buchhave}, {Ciardi},
  {Walkowicz}, {Sherry}, {Horch}, {Isaacson}, {Everett}, {Fischer}, {Torres},
  {Johnson}, {Endl}, {MacQueen}, {Bryson}, {Dotson}, {Haas}, {Kolodziejczak},
  {Van Cleve}, {Chandrasekaran}, {Twicken}, {Quintana}, {Clarke}, {Allen},
  {Li}, {Wu}, {Tenenbaum}, {Verner}, {Bruhweiler}, {Barnes}, \&
  {Prsa}}]{2010Sci...327..977B}
{Borucki}, W.~J., {Koch}, D., {Basri}, G., {et~al.} 2010, Science, 327, 977,
  \dodoi{10.1126/science.1185402}

\bibitem[{Brown \& Journaux(2020)}]{BrownJM2020}
Brown, J.~M., \& Journaux, B. 2020, Minerals, 10, \dodoi{10.3390/min10020092}

\bibitem[{{Brown} {et~al.}(2011){Brown}, {Latham}, {Everett}, \&
  {Esquerdo}}]{2011AJ....142..112B}
{Brown}, T.~M., {Latham}, D.~W., {Everett}, M.~E., \& {Esquerdo}, G.~A. 2011,
  \aj, 142, 112, \dodoi{10.1088/0004-6256/142/4/112}

\bibitem[{{Bryson} {et~al.}(2021){Bryson}, {Kunimoto}, {Kopparapu}, {Coughlin},
  {Borucki}, {Koch}, {Aguirre}, {Allen}, {Barentsen}, {Batalha}, {Berger},
  {Boss}, {Buchhave}, {Burke}, {Caldwell}, {Campbell}, {Catanzarite},
  {Chandrasekaran}, {Chaplin}, {Christiansen}, {Christensen-Dalsgaard},
  {Ciardi}, {Clarke}, {Cochran}, {Dotson}, {Doyle}, {Duarte}, {Dunham},
  {Dupree}, {Endl}, {Fanson}, {Ford}, {Fujieh}, {Gautier}, {Geary},
  {Gilliland}, {Girouard}, {Gould}, {Haas}, {Henze}, {Holman}, {Howard},
  {Howell}, {Huber}, {Hunter}, {Jenkins}, {Kjeldsen}, {Kolodziejczak},
  {Larson}, {Latham}, {Li}, {Mathur}, {Meibom}, {Middour}, {Morris}, {Morton},
  {Mullally}, {Mullally}, {Pletcher}, {Prsa}, {Quinn}, {Quintana}, {Ragozzine},
  {Ramirez}, {Sanderfer}, {Sasselov}, {Seader}, {Shabram}, {Shporer}, {Smith},
  {Steffen}, {Still}, {Torres}, {Troeltzsch}, {Twicken}, {Uddin}, {Van Cleve},
  {Voss}, {Weiss}, {Welsh}, {Wohler}, \& {Zamudio}}]{2021AJ....161...36B}
{Bryson}, S., {Kunimoto}, M., {Kopparapu}, R.~K., {et~al.} 2021, \aj, 161, 36,
  \dodoi{10.3847/1538-3881/abc418}

\bibitem[{{Chao} {et~al.}(2021){Chao}, {deGraffenried}, {Lach}, {Nelson},
  {Truax}, \& {Gaidos}}]{Chao2021}
{Chao}, K.-H., {deGraffenried}, R., {Lach}, M., {et~al.} 2021, Chemie der Erde
  / Geochemistry, 81, 125735, \dodoi{10.1016/j.chemer.2020.125735}

\bibitem[{{Claytor} {et~al.}(2020){Claytor}, {van Saders}, {Santos},
  {Garc{\'\i}a}, {Mathur}, {Tayar}, {Pinsonneault}, \&
  {Shetrone}}]{2020ascl.soft11027C}
{Claytor}, Z.~R., {van Saders}, J.~L., {Santos}, {\^A}. R.~G., {et~al.} 2020,
  {kiauhoku: Stellar model grid interpolation}.
\newblock \doeprint{2011.027}

\bibitem[{{Cottaar} {et~al.}(2016){Cottaar}, {Heister}, {Myhill}, {Rose}, \&
  {Unterborn}}]{2016ascl.soft10010C}
{Cottaar}, S., {Heister}, T., {Myhill}, R., {Rose}, I., \& {Unterborn}, C.
  2016, {BurnMan: Lower mantle mineral physics toolkit}.
\newblock \doeprint{1610.010}

\bibitem[{{Dai} {et~al.}(2019){Dai}, {Masuda}, {Winn}, \&
  {Zeng}}]{2019ApJ...883...79D}
{Dai}, F., {Masuda}, K., {Winn}, J.~N., \& {Zeng}, L. 2019, \apj, 883, 79,
  \dodoi{10.3847/1538-4357/ab3a3b}

\bibitem[{{Dewaele} {et~al.}(2006){Dewaele}, {Loubeyre}, {Occelli}, {Mezouar},
  {Dorogokupets}, \& {Torrent}}]{2006PhRvL..97u5504D}
{Dewaele}, A., {Loubeyre}, P., {Occelli}, F., {et~al.} 2006, \prl, 97, 215504,
  \dodoi{10.1103/PhysRevLett.97.215504}

\bibitem[{{Dressing} {et~al.}(2015){Dressing}, {Charbonneau}, {Dumusque},
  {Gettel}, {Pepe}, {Collier Cameron}, {Latham}, {Molinari}, {Udry}, {Affer},
  {Bonomo}, {Buchhave}, {Cosentino}, {Figueira}, {Fiorenzano}, {Harutyunyan},
  {Haywood}, {Johnson}, {Lopez-Morales}, {Lovis}, {Malavolta}, {Mayor},
  {Micela}, {Motalebi}, {Nascimbeni}, {Phillips}, {Piotto}, {Pollacco},
  {Queloz}, {Rice}, {Sasselov}, {S{\'e}gransan}, {Sozzetti}, {Szentgyorgyi}, \&
  {Watson}}]{2015ApJ...800..135D}
{Dressing}, C.~D., {Charbonneau}, D., {Dumusque}, X., {et~al.} 2015, \apj, 800,
  135, \dodoi{10.1088/0004-637X/800/2/135}

\bibitem[{{Dumusque} {et~al.}(2011){Dumusque}, {Udry}, {Lovis}, {Santos}, \&
  {Monteiro}}]{2011A&A...525A.140D}
{Dumusque}, X., {Udry}, S., {Lovis}, C., {Santos}, N.~C., \& {Monteiro},
  M.~J.~P.~F.~G. 2011, \aap, 525, A140, \dodoi{10.1051/0004-6361/201014097}

\bibitem[{{Dumusque} {et~al.}(2014){Dumusque}, {Bonomo}, {Haywood},
  {Malavolta}, {S{\'e}gransan}, {Buchhave}, {Collier Cameron}, {Latham},
  {Molinari}, {Pepe}, {Udry}, {Charbonneau}, {Cosentino}, {Dressing},
  {Figueira}, {Fiorenzano}, {Gettel}, {Harutyunyan}, {Horne}, {Lopez-Morales},
  {Lovis}, {Mayor}, {Micela}, {Motalebi}, {Nascimbeni}, {Phillips}, {Piotto},
  {Pollacco}, {Queloz}, {Rice}, {Sasselov}, {Sozzetti}, {Szentgyorgyi}, \&
  {Watson}}]{2014ApJ...789..154D}
{Dumusque}, X., {Bonomo}, A.~S., {Haywood}, R.~D., {et~al.} 2014, \apj, 789,
  154, \dodoi{10.1088/0004-637X/789/2/154}

\bibitem[{{Dumusque} {et~al.}(2021){Dumusque}, {Cretignier}, {Sosnowska},
  {Buchschacher}, {Lovis}, {Phillips}, {Pepe}, {Alesina}, {Buchhave},
  {Burnier}, {Cecconi}, {Cegla}, {Cloutier}, {Collier Cameron}, {Cosentino},
  {Ghedina}, {Gonz{\'a}lez}, {Haywood}, {Latham}, {Lodi}, {L{\'o}pez-Morales},
  {Maldonado}, {Malavolta}, {Micela}, {Molinari}, {Mortier}, {P{\'e}rez
  Ventura}, {Pinamonti}, {Poretti}, {Rice}, {Riverol}, {Riverol}, {San Juan},
  {S{\'e}gransan}, {Sozzetti}, {Thompson}, {Udry}, \&
  {Wilson}}]{2021A&A...648A.103D}
{Dumusque}, X., {Cretignier}, M., {Sosnowska}, D., {et~al.} 2021, \aap, 648,
  A103, \dodoi{10.1051/0004-6361/202039350}

\bibitem[{{Dziewonski} \& {Anderson}(1981)}]{1981PEPI...25..297D}
{Dziewonski}, A.~M., \& {Anderson}, D.~L. 1981, Physics of the Earth and
  Planetary Interiors, 25, 297, \dodoi{10.1016/0031-9201(81)90046-7}

\bibitem[{{Eastman} {et~al.}(2013){Eastman}, {Gaudi}, \&
  {Agol}}]{2013PASP..125...83E}
{Eastman}, J., {Gaudi}, B.~S., \& {Agol}, E. 2013, \pasp, 125, 83,
  \dodoi{10.1086/669497}

\bibitem[{{Elkins-Tanton} \& {Seager}(2008)}]{2008ApJ...688..628E}
{Elkins-Tanton}, L.~T., \& {Seager}, S. 2008, \apj, 688, 628,
  \dodoi{10.1086/592316}

\bibitem[{{Foreman-Mackey} {et~al.}(2013){Foreman-Mackey}, {Hogg}, {Lang}, \&
  {Goodman}}]{2013PASP..125..306F}
{Foreman-Mackey}, D., {Hogg}, D.~W., {Lang}, D., \& {Goodman}, J. 2013, \pasp,
  125, 306, \dodoi{10.1086/670067}

\bibitem[{Frank {et~al.}(2004)Frank, Fei, \& Hu}]{FRANK20042781}
Frank, M.~R., Fei, Y., \& Hu, J. 2004, Geochimica et Cosmochimica Acta, 68,
  2781, \dodoi{https://doi.org/10.1016/j.gca.2003.12.007}

\bibitem[{{Fulton} \& {Petigura}(2018)}]{2018AJ....156..264F}
{Fulton}, B.~J., \& {Petigura}, E.~A. 2018, \aj, 156, 264,
  \dodoi{10.3847/1538-3881/aae828}

\bibitem[{{Fulton} {et~al.}(2018){Fulton}, {Petigura}, {Blunt}, \&
  {Sinukoff}}]{2018PASP..130d4504F}
{Fulton}, B.~J., {Petigura}, E.~A., {Blunt}, S., \& {Sinukoff}, E. 2018, \pasp,
  130, 044504, \dodoi{10.1088/1538-3873/aaaaa8}

\bibitem[{{Fulton} {et~al.}(2017){Fulton}, {Petigura}, {Howard}, {Isaacson},
  {Marcy}, {Cargile}, {Hebb}, {Weiss}, {Johnson}, {Morton}, {Sinukoff},
  {Crossfield}, \& {Hirsch}}]{2017AJ....154..109F}
{Fulton}, B.~J., {Petigura}, E.~A., {Howard}, A.~W., {et~al.} 2017, \aj, 154,
  109, \dodoi{10.3847/1538-3881/aa80eb}

\bibitem[{{Gajdo{\v{s}}} {et~al.}(2019){Gajdo{\v{s}}}, {Va{\v{n}}ko}, \&
  {Parimucha}}]{2019RAA....19...41G}
{Gajdo{\v{s}}}, P., {Va{\v{n}}ko}, M., \& {Parimucha}, {\v{S}}. 2019, Research
  in Astronomy and Astrophysics, 19, 041, \dodoi{10.1088/1674-4527/19/3/41}

\bibitem[{{Gibson} {et~al.}(2012){Gibson}, {Aigrain}, {Roberts}, {Evans},
  {Osborne}, \& {Pont}}]{2012MNRAS.419.2683G}
{Gibson}, N.~P., {Aigrain}, S., {Roberts}, S., {et~al.} 2012, \mnras, 419,
  2683, \dodoi{10.1111/j.1365-2966.2011.19915.x}

\bibitem[{{Grunblatt} {et~al.}(2016){Grunblatt}, {Howard}, \&
  {Haywood}}]{2016IAUFM..29A.208G}
{Grunblatt}, S.~K., {Howard}, A.~W., \& {Haywood}, R.~D. 2016, IAU Focus
  Meeting, 29A, 208, \dodoi{10.1017/S1743921316002829}

\bibitem[{Harris {et~al.}(2020)Harris, Millman, van~der Walt, Gommers,
  Virtanen, Cournapeau, Wieser, Taylor, Berg, Smith, Kern, Picus, Hoyer, van
  Kerkwijk, Brett, Haldane, del R{\'{i}}o, Wiebe, Peterson,
  G{\'{e}}rard-Marchant, Sheppard, Reddy, Weckesser, Abbasi, Gohlke, \&
  Oliphant}]{harris2020array}
Harris, C.~R., Millman, K.~J., van~der Walt, S.~J., {et~al.} 2020, Nature, 585,
  357, \dodoi{10.1038/s41586-020-2649-2}

\bibitem[{{Haywood} {et~al.}(2014){Haywood}, {Collier Cameron}, {Queloz},
  {Barros}, {Deleuil}, {Fares}, {Gillon}, {Lanza}, {Lovis}, {Moutou}, {Pepe},
  {Pollacco}, {Santerne}, {S{\'e}gransan}, \& {Unruh}}]{2014MNRAS.443.2517H}
{Haywood}, R.~D., {Collier Cameron}, A., {Queloz}, D., {et~al.} 2014, \mnras,
  443, 2517, \dodoi{10.1093/mnras/stu1320}

\bibitem[{{Howard} {et~al.}(2010){Howard}, {Johnson}, {Marcy}, {Fischer},
  {Wright}, {Bernat}, {Henry}, {Peek}, {Isaacson}, {Apps}, {Endl}, {Cochran},
  {Valenti}, {Anderson}, \& {Piskunov}}]{2010ApJ...721.1467H}
{Howard}, A.~W., {Johnson}, J.~A., {Marcy}, G.~W., {et~al.} 2010, \apj, 721,
  1467, \dodoi{10.1088/0004-637X/721/2/1467}

\bibitem[{{Howard} {et~al.}(2012){Howard}, {Marcy}, {Bryson}, {Jenkins},
  {Rowe}, {Batalha}, {Borucki}, {Koch}, {Dunham}, {Gautier}, {Van Cleve},
  {Cochran}, {Latham}, {Lissauer}, {Torres}, {Brown}, {Gilliland}, {Buchhave},
  {Caldwell}, {Christensen-Dalsgaard}, {Ciardi}, {Fressin}, {Haas}, {Howell},
  {Kjeldsen}, {Seager}, {Rogers}, {Sasselov}, {Steffen}, {Basri},
  {Charbonneau}, {Christiansen}, {Clarke}, {Dupree}, {Fabrycky}, {Fischer},
  {Ford}, {Fortney}, {Tarter}, {Girouard}, {Holman}, {Johnson}, {Klaus},
  {Machalek}, {Moorhead}, {Morehead}, {Ragozzine}, {Tenenbaum}, {Twicken},
  {Quinn}, {Isaacson}, {Shporer}, {Lucas}, {Walkowicz}, {Welsh}, {Boss},
  {Devore}, {Gould}, {Smith}, {Morris}, {Prsa}, {Morton}, {Still}, {Thompson},
  {Mullally}, {Endl}, \& {MacQueen}}]{2012ApJS..201...15H}
{Howard}, A.~W., {Marcy}, G.~W., {Bryson}, S.~T., {et~al.} 2012, \apjs, 201,
  15, \dodoi{10.1088/0067-0049/201/2/15}

\bibitem[{Hunter(2007)}]{Hunter:2007}
Hunter, J.~D. 2007, Computing in Science \& Engineering, 9, 90,
  \dodoi{10.1109/MCSE.2007.55}

\bibitem[{{Journaux} {et~al.}(2019){Journaux}, {Brown}, {Pakhomova},
  {Collings}, {Petitgirard}, {Espinoza}, {Ott}, {Cova}, {Garbarino}, \&
  {Hanfland}}]{2019arXiv190709598J}
{Journaux}, B., {Brown}, J.~M., {Pakhomova}, A., {et~al.} 2019, arXiv e-prints,
  arXiv:1907.09598.
\newblock \doarXiv{1907.09598}

\bibitem[{Kass \& Raftery(1995)}]{doi:10.1080/01621459.1995.10476572}
Kass, R.~E., \& Raftery, A.~E. 1995, Journal of the American Statistical
  Association, 90, 773, \dodoi{10.1080/01621459.1995.10476572}

\bibitem[{{Kite} {et~al.}(2020){Kite}, {Fegley}, {Schaefer}, \&
  {Ford}}]{2020ApJ...891..111K}
{Kite}, E.~S., {Fegley}, Bruce, J., {Schaefer}, L., \& {Ford}, E.~B. 2020,
  \apj, 891, 111, \dodoi{10.3847/1538-4357/ab6ffb}

\bibitem[{{Kosiarek} {et~al.}(2019){Kosiarek}, {Blunt}, {L{\'o}pez-Morales},
  {Crossfield}, {Sinukoff}, {Petigura}, {Gonzales}, {Poretti}, {Malavolta},
  {Howard}, {Isaacson}, {Haywood}, {Ciardi}, {Bristow}, {Collier Cameron},
  {Charbonneau}, {Dressing}, {Figueira}, {Fulton}, {Hardee}, {Hirsch},
  {Latham}, {Mortier}, {Nava}, {Schlieder}, {Vanderburg}, {Weiss}, {Bonomo},
  {Bouchy}, {Buchhave}, {Coffinet}, {Damasso}, {Dumusque}, {Lovis}, {Mayor},
  {Micela}, {Molinari}, {Pepe}, {Phillips}, {Piotto}, {Rice}, {Sasselov},
  {S{\'e}gransan}, {Sozzetti}, {Udry}, \& {Watson}}]{2019AJ....157..116K}
{Kosiarek}, M.~R., {Blunt}, S., {L{\'o}pez-Morales}, M., {et~al.} 2019, \aj,
  157, 116, \dodoi{10.3847/1538-3881/aafe83}

\bibitem[{{Lightkurve Collaboration} {et~al.}(2018){Lightkurve Collaboration},
  {Cardoso}, {Hedges}, {Gully-Santiago}, {Saunders}, {Cody}, {Barclay}, {Hall},
  {Sagear}, {Turtelboom}, {Zhang}, {Tzanidakis}, {Mighell}, {Coughlin}, {Bell},
  {Berta-Thompson}, {Williams}, {Dotson}, \& {Barentsen}}]{2018ascl.soft12013L}
{Lightkurve Collaboration}, {Cardoso}, J. V. d.~M., {Hedges}, C., {et~al.}
  2018, {Lightkurve: Kepler and TESS time series analysis in Python}.
\newblock \doeprint{1812.013}

\bibitem[{{Liu} {et~al.}(2007){Liu}, {San Liang}, \&
  {Weisberg}}]{2007JAtOT..24.2093L}
{Liu}, Y., {San Liang}, X., \& {Weisberg}, R.~H. 2007, Journal of Atmospheric
  and Oceanic Technology, 24, 2093, \dodoi{10.1175/2007JTECHO511.1}

\bibitem[{{Lodders}(2003)}]{2003ApJ...591.1220L}
{Lodders}, K. 2003, \apj, 591, 1220, \dodoi{10.1086/375492}

\bibitem[{{Lopez}(2017)}]{2017MNRAS.472..245L}
{Lopez}, E.~D. 2017, \mnras, 472, 245, \dodoi{10.1093/mnras/stx1558}

\bibitem[{{Lopez} \& {Fortney}(2014)}]{2014ApJ...792....1L}
{Lopez}, E.~D., \& {Fortney}, J.~J. 2014, \apj, 792, 1,
  \dodoi{10.1088/0004-637X/792/1/1}

\bibitem[{{L{\'o}pez-Morales} \& {Seager}(2007)}]{2007ApJ...667L.191L}
{L{\'o}pez-Morales}, M., \& {Seager}, S. 2007, \apjl, 667, L191,
  \dodoi{10.1086/522118}

\bibitem[{Lorah \& Womack(2019)}]{2019Lorah}
Lorah, J., \& Womack, A. 2019, Behavior Research Methods, 51, 440,
  \dodoi{10.3758/s13428-018-1188-3}

\bibitem[{{Luger} \& {Barnes}(2015)}]{2015AsBio..15..119L}
{Luger}, R., \& {Barnes}, R. 2015, Astrobiology, 15, 119,
  \dodoi{10.1089/ast.2014.1231}

\bibitem[{{Malavolta} {et~al.}(2016){Malavolta}, {Nascimbeni}, {Piotto},
  {Quinn}, {Borsato}, {Granata}, {Bonomo}, {Marzari}, {Bedin}, {Rainer},
  {Desidera}, {Lanza}, {Poretti}, {Sozzetti}, {White}, {Latham}, {Cunial},
  {Libralato}, {Nardiello}, {Boccato}, {Claudi}, {Cosentino}, {Covino},
  {Gratton}, {Maggio}, {Micela}, {Molinari}, {Pagano}, {Smareglia}, {Affer},
  {Andreuzzi}, {Aparicio}, {Benatti}, {Bignamini}, {Borsa}, {Damasso}, {Di
  Fabrizio}, {Harutyunyan}, {Esposito}, {Fiorenzano}, {Gandolfi}, {Giacobbe},
  {Gonz{\'a}lez Hern{\'a}ndez}, {Maldonado}, {Masiero}, {Molinaro}, {Pedani},
  \& {Scandariato}}]{2016A&A...588A.118M}
{Malavolta}, L., {Nascimbeni}, V., {Piotto}, G., {et~al.} 2016, \aap, 588,
  A118, \dodoi{10.1051/0004-6361/201527933}

\bibitem[{{Malavolta} {et~al.}(2018){Malavolta}, {Mayo}, {Louden}, {Rajpaul},
  {Bonomo}, {Buchhave}, {Kreidberg}, {Kristiansen}, {Lopez-Morales}, {Mortier},
  {Vanderburg}, {Coffinet}, {Ehrenreich}, {Lovis}, {Bouchy}, {Charbonneau},
  {Ciardi}, {Collier Cameron}, {Cosentino}, {Crossfield}, {Damasso},
  {Dressing}, {Dumusque}, {Everett}, {Figueira}, {Fiorenzano}, {Gonzales},
  {Haywood}, {Harutyunyan}, {Hirsch}, {Howell}, {Johnson}, {Latham}, {Lopez},
  {Mayor}, {Micela}, {Molinari}, {Nascimbeni}, {Pepe}, {Phillips}, {Piotto},
  {Rice}, {Sasselov}, {S{\'e}gransan}, {Sozzetti}, {Udry}, \&
  {Watson}}]{2018AJ....155..107M}
{Malavolta}, L., {Mayo}, A.~W., {Louden}, T., {et~al.} 2018, \aj, 155, 107,
  \dodoi{10.3847/1538-3881/aaa5b5}

\bibitem[{{Marcy} \& {Butler}(1992)}]{1992PASP..104..270M}
{Marcy}, G.~W., \& {Butler}, R.~P. 1992, \pasp, 104, 270,
  \dodoi{10.1086/132989}

\bibitem[{{Marcy} {et~al.}(2014){Marcy}, {Isaacson}, {Howard}, {Rowe},
  {Jenkins}, {Bryson}, {Latham}, {Howell}, {Gautier}, {Batalha}, {Rogers},
  {Ciardi}, {Fischer}, {Gilliland}, {Kjeldsen}, {Christensen-Dalsgaard},
  {Huber}, {Chaplin}, {Basu}, {Buchhave}, {Quinn}, {Borucki}, {Koch}, {Hunter},
  {Caldwell}, {Van Cleve}, {Kolbl}, {Weiss}, {Petigura}, {Seager}, {Morton},
  {Johnson}, {Ballard}, {Burke}, {Cochran}, {Endl}, {MacQueen}, {Everett},
  {Lissauer}, {Ford}, {Torres}, {Fressin}, {Brown}, {Steffen}, {Charbonneau},
  {Basri}, {Sasselov}, {Winn}, {Sanchis-Ojeda}, {Christiansen}, {Adams},
  {Henze}, {Dupree}, {Fabrycky}, {Fortney}, {Tarter}, {Holman}, {Tenenbaum},
  {Shporer}, {Lucas}, {Welsh}, {Orosz}, {Bedding}, {Campante}, {Davies},
  {Elsworth}, {Handberg}, {Hekker}, {Karoff}, {Kawaler}, {Lund}, {Lundkvist},
  {Metcalfe}, {Miglio}, {Silva Aguirre}, {Stello}, {White}, {Boss}, {Devore},
  {Gould}, {Prsa}, {Agol}, {Barclay}, {Coughlin}, {Brugamyer}, {Mullally},
  {Quintana}, {Still}, {Thompson}, {Morrison}, {Twicken}, {D{\'e}sert},
  {Carter}, {Crepp}, {H{\'e}brard}, {Santerne}, {Moutou}, {Sobeck}, {Hudgins},
  {Haas}, {Robertson}, {Lillo-Box}, \& {Barrado}}]{2014ApJS..210...20M}
{Marcy}, G.~W., {Isaacson}, H., {Howard}, A.~W., {et~al.} 2014, \apjs, 210, 20,
  \dodoi{10.1088/0067-0049/210/2/20}

\bibitem[{{Morton} {et~al.}(2016){Morton}, {Bryson}, {Coughlin}, {Rowe},
  {Ravichandran}, {Petigura}, {Haas}, \& {Batalha}}]{2016ApJ...822...86M}
{Morton}, T.~D., {Bryson}, S.~T., {Coughlin}, J.~L., {et~al.} 2016, \apj, 822,
  86, \dodoi{10.3847/0004-637X/822/2/86}

\bibitem[{{Otegi} {et~al.}(2020){Otegi}, {Bouchy}, \&
  {Helled}}]{2020A&A...634A..43O}
{Otegi}, J.~F., {Bouchy}, F., \& {Helled}, R. 2020, \aap, 634, A43,
  \dodoi{10.1051/0004-6361/201936482}

\bibitem[{{Owen} \& {Wu}(2017)}]{2017ApJ...847...29O}
{Owen}, J.~E., \& {Wu}, Y. 2017, \apj, 847, 29,
  \dodoi{10.3847/1538-4357/aa890a}

\bibitem[{{Petigura} {et~al.}(2013){Petigura}, {Howard}, \&
  {Marcy}}]{2013PNAS..11019273P}
{Petigura}, E.~A., {Howard}, A.~W., \& {Marcy}, G.~W. 2013, Proceedings of the
  National Academy of Science, 110, 19273, \dodoi{10.1073/pnas.1319909110}

\bibitem[{{Rasmussen} \& {Williams}(2006)}]{2006gpml.book.....R}
{Rasmussen}, C.~E., \& {Williams}, C. K.~I. 2006, {Gaussian Processes for
  Machine Learning}

\bibitem[{{Rogers}(2015)}]{2015ApJ...801...41R}
{Rogers}, L.~A. 2015, \apj, 801, 41, \dodoi{10.1088/0004-637X/801/1/41}

\bibitem[{{Rogers} \& {Seager}(2010)}]{2010ApJ...712..974R}
{Rogers}, L.~A., \& {Seager}, S. 2010, \apj, 712, 974,
  \dodoi{10.1088/0004-637X/712/2/974}

\bibitem[{{Schr{\"o}der} {et~al.}(2013){Schr{\"o}der}, {Mittag}, {Hempelmann},
  {Gonz{\'a}lez-P{\'e}rez}, \& {Schmitt}}]{2013A&A...554A..50S}
{Schr{\"o}der}, K.~P., {Mittag}, M., {Hempelmann}, A.,
  {Gonz{\'a}lez-P{\'e}rez}, J.~N., \& {Schmitt}, J.~H.~M.~M. 2013, \aap, 554,
  A50, \dodoi{10.1051/0004-6361/201219830}

\bibitem[{{Stixrude} \& {Lithgow-Bertelloni}(2011)}]{2011GeoJI.184.1180S}
{Stixrude}, L., \& {Lithgow-Bertelloni}, C. 2011, Geophysical Journal
  International, 184, 1180, \dodoi{10.1111/j.1365-246X.2010.04890.x}

\bibitem[{{Valencia} {et~al.}(2007){Valencia}, {Sasselov}, \&
  {O'Connell}}]{2007ApJ...656..545V}
{Valencia}, D., {Sasselov}, D.~D., \& {O'Connell}, R.~J. 2007, \apj, 656, 545,
  \dodoi{10.1086/509800}

\bibitem[{{van Saders} {et~al.}(2016){van Saders}, {Ceillier}, {Metcalfe},
  {Silva Aguirre}, {Pinsonneault}, {Garc{\'\i}a}, {Mathur}, \&
  {Davies}}]{2016Natur.529..181V}
{van Saders}, J.~L., {Ceillier}, T., {Metcalfe}, T.~S., {et~al.} 2016, \nat,
  529, 181, \dodoi{10.1038/nature16168}

\bibitem[{{van Saders} \& {Pinsonneault}(2013)}]{2013ApJ...776...67V}
{van Saders}, J.~L., \& {Pinsonneault}, M.~H. 2013, \apj, 776, 67,
  \dodoi{10.1088/0004-637X/776/2/67}

\bibitem[{{Virtanen} {et~al.}(2020){Virtanen}, {Gommers}, {Burovski},
  {Oliphant}, {Weckesser}, {Cournapeau}, {Alexbrc}, {Peterson}, {Reddy},
  {Wilson}, {Haberland}, {Mayorov}, {Endolith}, {Nelson}, {Van Der Walt},
  {Laxalde}, {Brett}, {Polat}, {Larson}, {Millman}, {Lars}, {Van Mulbregt},
  {Eric-Jones}, {Carey}, {Moore}, {Kern}, {Leslie}, {Perktold}, {Striega}, \&
  {Feng}}]{2020zndo...4100507V}
{Virtanen}, P., {Gommers}, R., {Burovski}, E., {et~al.} 2020, {scipy/scipy:
  SciPy 1.5.3}, v1.5.3,  Zenodo, \dodoi{10.5281/zenodo.4100507}

\bibitem[{Virtanen {et~al.}(2020)Virtanen, Gommers, Oliphant, Haberland, Reddy,
  Cournapeau, Burovski, Peterson, Weckesser, Bright, {van der Walt}, Brett,
  Wilson, Millman, Mayorov, Nelson, Jones, Kern, Larson, Carey, Polat, Feng,
  Moore, {VanderPlas}, Laxalde, Perktold, Cimrman, Henriksen, Quintero, Harris,
  Archibald, Ribeiro, Pedregosa, {van Mulbregt}, \& {SciPy 1.0
  Contributors}}]{2020SciPy-NMeth}
Virtanen, P., Gommers, R., Oliphant, T.~E., {et~al.} 2020, Nature Methods, 17,
  261, \dodoi{10.1038/s41592-019-0686-2}

\bibitem[{{Vogt} {et~al.}(1994){Vogt}, {Allen}, {Bigelow}, {Bresee}, {Brown},
  {Cantrall}, {Conrad}, {Couture}, {Delaney}, {Epps}, {Hilyard}, {Hilyard},
  {Horn}, {Jern}, {Kanto}, {Keane}, {Kibrick}, {Lewis}, {Osborne},
  {Pardeilhan}, {Pfister}, {Ricketts}, {Robinson}, {Stover}, {Tucker}, {Ward},
  \& {Wei}}]{1994SPIE.2198..362V}
{Vogt}, S.~S., {Allen}, S.~L., {Bigelow}, B.~C., {et~al.} 1994, in Society of
  Photo-Optical Instrumentation Engineers (SPIE) Conference Series, Vol. 2198,
  Instrumentation in Astronomy VIII, ed. D.~L. {Crawford} \& E.~R. {Craine},
  362, \dodoi{10.1117/12.176725}

\bibitem[{{Wagner} \& {Pru{\ss}}(2002)}]{2002JPCRD..31..387W}
{Wagner}, W., \& {Pru{\ss}}, A. 2002, Journal of Physical and Chemical
  Reference Data, 31, 387, \dodoi{10.1063/1.1461829}

\bibitem[{{Weiss} \& {Marcy}(2014)}]{2014ApJ...783L...6W}
{Weiss}, L.~M., \& {Marcy}, G.~W. 2014, \apjl, 783, L6,
  \dodoi{10.1088/2041-8205/783/1/L6}

\bibitem[{{Weiss} {et~al.}(2016){Weiss}, {Rogers}, {Isaacson}, {Agol}, {Marcy},
  {Rowe}, {Kipping}, {Fulton}, {Lissauer}, {Howard}, \&
  {Fabrycky}}]{2016ApJ...819...83W}
{Weiss}, L.~M., {Rogers}, L.~A., {Isaacson}, H.~T., {et~al.} 2016, \apj, 819,
  83, \dodoi{10.3847/0004-637X/819/1/83}

\bibitem[{{W}es {M}c{K}inney(2010)}]{mckinney-proc-scipy-2010}
{W}es {M}c{K}inney. 2010, in {P}roceedings of the 9th {P}ython in {S}cience
  {C}onference, ed. {S}t\'efan van~der {W}alt \& {J}arrod {M}illman, 56 -- 61,
  \dodoi{10.25080/Majora-92bf1922-00a}

\bibitem[{{Yee} {et~al.}(2021){Yee}, {Tamayo}, {Hadden}, \&
  {Winn}}]{2021AJ....162...55Y}
{Yee}, S.~W., {Tamayo}, D., {Hadden}, S., \& {Winn}, J.~N. 2021, \aj, 162, 55,
  \dodoi{10.3847/1538-3881/ac00a9}

\bibitem[{{Zeng} {et~al.}(2019){Zeng}, {Jacobsen}, {Sasselov}, {Petaev},
  {Vanderburg}, {Lopez-Morales}, {Perez-Mercader}, {Mattsson}, {Li}, {Heising},
  {Bonomo}, {Damasso}, {Berger}, {Cao}, {Levi}, \&
  {Wordsworth}}]{2019PNAS..116.9723Z}
{Zeng}, L., {Jacobsen}, S.~B., {Sasselov}, D.~D., {et~al.} 2019, Proceedings of
  the National Academy of Science, 116, 9723, \dodoi{10.1073/pnas.1812905116}

\end{thebibliography}
\bibliographystyle{aasjournal}

\end{document}